\documentclass[prb,aps,twocolumn,showpacs,superscriptaddress]{revtex4}

\usepackage{amsmath, amsfonts, amssymb}
\usepackage{enumerate}
\usepackage{multirow}
\usepackage{graphicx}
\usepackage{color}

\newcommand{\be} 		{\begin{equation}}
\newcommand{\ee} 		{\end{equation}}
\newcommand{\ba} 		{\begin{eqnarray}}
\newcommand{\ea} 		{\end{eqnarray}}
\newcommand{\up} 		{\uparrow}
\newcommand{\dw} 		{\downarrow}
\newcommand{\e}	 		{\mathrm{e}}
\newcommand{\p}  		{\partial}
\newcommand{\s}  		{\sigma}

\newcommand{\sign}		{\mathrm{sign}}
\newcommand{\mean}[1]	{\langle #1 \rangle}

\begin{document}


\title{ 
Spectral Properties of Luttinger Liquids:
A Comparative Analysis of Regular, Helical, and Spiral Luttinger Liquids
}

\author{Bernd Braunecker}
\affiliation{Departamento de F{\'\i}sica Te\'{o}rica de la Materia Condensada, Facultad de Ciencias, 
             Universidad Aut\'{o}noma de Madrid, 28049 Madrid, Spain}
\affiliation{Department of Physics, University of Basel, Klingelbergstrasse 82, 4056 Basel, Switzerland}
\author{Cristina Bena}
\affiliation{Laboratoire de Physique des Solides, CNRS UMR-8502, Universit\'{e} Paris Sud, 91405 Orsay Cedex, France}
\affiliation{Institute de Physique Th\a'eorique, CEA/Saclay, Orme des Merisiers, 91190 Gif-sur-Yvette Cedex, France}
\author{Pascal Simon}
\affiliation{Laboratoire de Physique des Solides, CNRS UMR-8502, Universit\'{e} Paris Sud, 91405 Orsay Cedex, France}

\date{\today}


\begin{abstract}
We provide analytic expressions for the Green's functions in position-frequency space as well as for the 
tunneling density of states of various Luttinger liquids at zero temperature:
the standard spinless and spinful Luttinger liquids, the helical Luttinger liquid
at the edge of a topological insulator, and the Luttinger liquid that appears either together
with an ordering transition of nuclear spins in a one-dimensional conductor, or in spin-orbit
split quantum wires in an external magnetic field. The latter system is often used to mimic a helical 
Luttinger liquid, yet we show here that it exhibits significantly different response functions
and, to discriminate, we call it the spiral Luttinger liquid.
We give fully analytic results for the tunneling density of state of all the Luttinger liquids 
as well as for most of the Green's functions. The remaining Green's functions
are expressed by simple convolution integrals between analytic results.
\end{abstract}




\maketitle


\section{Introduction}

Helical conductors have received much attention recently. They consist of one-dimensional (1D) 
conductors exhibiting spin-filtered transport, for instance, of right-moving modes carrying
spin up, $(R \uparrow)$, and left-moving modes carrying spin down, $(L \downarrow)$.
Such conductors appear on the edges of two-dimensional (2D) topological insulators
(where spin should more correctly be addressed as Kramers partner),\cite{qshe,hasan:2010}
in nanowires made of tree-dimensional (3D) topological insulators,\cite{egger:2010,kong:2010,peng:2010}
in 1D semiconductor wires with strong spin-orbit interaction in an external magnetic field
(with approximate spin-filtering),\cite{streda:2003,pershin:2004,devillard:2005,zhang:2006,sanchez:2008,birkholz:2009,quay:2010,bjkl}
or in carbon nanotubes, where the combination of spin-orbit interaction
and strong external electric fields can cause helical conduction.\cite{ksbl1,ksbl2}
Perfect spin-filtered conduction modes can also appear in regular 1D conductors in the 
presence of nuclear spins as explained in further detail below.
Aside from the spin-filtering, if a helical conductor is brought in the proximity of a superconductor
it supports Majorana bound states at its ends.\cite{fu:2008,alicea:2010,sau:2010a,sau:2010b,tewari:2010,lutchyn:2010,oreg:2010,alicea:2011,klinovaja:2012}
Such Majorana end states have attracted much interest recently as parts of elementary 
excitations with nonabelian statistics that may be useful for topological
quantum computation.\cite{kitaev:2001,nayak:2008,hassler:2010,wimmer:2010,alicea:2011,duckheim:2011}

While much of the discussion of these systems in the literature is focused on the fundamental physics 
and the possible applications of the helical conductors, it is still a challenge
to provide a unique experimental proof of the existence of the helical states. 
In this paper we show that probing the spectral properties of 1D conductors
provides an extra tool to investigate experimentally the differences between helical and non-helical systems.
This complements recent proposals to probe spin-dependent transport properties in helical conductors 
using, for instance, a quantum-point-contact setup or a Scanning Tunneling Microscope (STM).\cite{hou:2009,strom:2009,teo:2009,das:2011}

Using the Luttinger liquid (LL) theory, which allows us to take into account the effects of the 
electron-electron interactions, we compute the position-frequency dependent Green's functions 
for the helical conductors proposed in the literature and, for comparison, for the regular spinless 
and spinful LLs. 
This allows us to obtain information about the spectral properties of these systems.
In particular, we give explicit expressions for the local (or tunneling) density of states (DOS) 
that may be probed directly, for instance, by STM experiments. 
Most of our results are fully analytic which, to our knowledge,
has not been achieved before, even for regular LLs. We show that electron-electron interactions
play a central role and determine much of the spectral properties. 
In particular, we show that the interactions cause a substantially
different response of the conduction modes in 1D quantum wires as compared with 
the conduction modes at the edge of a topological insulator.

Furthermore, our results provide a handle to detect nuclear spin order in 
a 1D conductor. As mentioned above, it has recently been shown that the presence of nuclear 
spins in an electron liquid can lead to new physics at low-enough 
temperatures. In a semiconductor, such as GaAs,
nuclear spins and electrons interact via the hyperfine interaction. This generates a long-ranged interaction among
the nuclear spins mediated by the electron gas, which is of  Ruderman-Kittel-Kasuya-Yosida (RKKY) type.
In 2D and 1D conductors, electron-electron interactions increase the
long-range nature of the RKKY interaction (for the 1D case, see Ref. \onlinecite{egger:1996}) and can trigger a finite temperature ordering of the nuclear 
spins in systems of finite size.\cite{simon07,simon08,bsl1,bsl2,loss11} 
The nuclear magnetic ordering has only little effect on the electron state in 2D,\cite{simon07,simon08}
which may hinder a possible detection of this order via electronic transport or spectroscopy.

The situation differs in 1D metallic systems, where electron-electron interactions have a dramatic effect.
The Fermi liquid description breaks down and the physics is dominated by collective low energy excitations.
The appropriate description of this scenario is known as LL theory
(see, for instance, Refs. \onlinecite{gogolin} and \onlinecite{giamarchi} for reviews).
Signatures of the LL physics have been observed nowadays in many systems such as
in semiconducting GaAs quantum wires,\cite{yacoby02,yacoby05,yacoby08,jompol09}
in metallic single wall carbon nanotubes,\cite{bockrath,dekker,bachtold,ihara} 
in fractional quantum Hall edge states,\cite{chang96}
in bundles of NbSe$_3$ nanowires,\cite{slot04}
in polymer nanofibers,\cite{aleshin04}
in MoSe nanowires,\cite{venkataraman06}
in conjugated polymers at high carrier densities,\cite{yuen09}
and
in atomically controlled chains of gold atoms on Ge surfaces.\cite{blumenstein:2011}

On a spinoff from the LL theory, it has been shown in Refs. \onlinecite{bsl1} and \onlinecite{bsl2} that in 
the presence of a lattice of nuclear moments and of the
hyperfine interaction, an exotic ordered phase emerges at low temperature in which the magnetic moments of the nuclei 
and the conduction
electrons of a 1D conductor are bound together due to the RKKY interaction. 
The nuclear spins order in form of a helimagnet. In turn, the resulting magnetic Overhauser
field strongly affects the conduction electrons, much in contrast to the 2D case. 
It induces a gap for one half of the low-energy modes, which
yields a partial electron spin polarization that follows the nuclear spiral polarization. The rest of the
low-energy modes remains gapless and can be described by an effective LL model with perfectly spin-filtered
left- and right-moving modes. We call this LL state a ``spiral Luttinger liquid'' (SLL). 
A similar SLL model has been shown to describe the quantum wires with spin-orbit interaction in the presence of
an external magnetic field, in which the spin-dependent band shift induced by the spin-orbit interaction
is commensurate with the Fermi momentum $k_F$.\cite{bjkl}
The latter type of LL corresponds thus to a special point in the general LL 
description of interacting quantum wires with spin-orbit 
interaction.\cite{moroz:2000a,moroz:2000b,governale:2002,iucci:2003,yu:2004,gritsev:2005,cheng:2007,
sun:2007,gangadharaiah:2008,schulz:2009,japaridze:2009,schulz:2010,stroem:2010,malard:2011}

The SLL should not be confused with the helical Luttinger liquid (HLL), which has been introduced 
to describe the edge states of 2D topological insulators.\cite{wu06} 
The HLL is also spin-filtered, 
but differs from the SLL, the most important difference being that the SLL describes the bulk properties of a 1D system, 
while the HLL describes the edge of a 2D system, or the surface of a nanowire made of a 3D topological insulator.\cite{egger:2010}
For noninteracting electrons, the low-energy physics of HLLs and SLLs is 
equivalent and, as mentioned, this equivalence has been of much interest very recently
because both HLLs and SLLs support the Majorana bound states at their ends
if they are brought in the proximity of a 
superconductor.\cite{fu:2008,alicea:2010,sau:2010a,sau:2010b,tewari:2010,lutchyn:2010,oreg:2010,alicea:2011,klinovaja:2012}
For interacting electrons, the proximity induced superconductivity and the 
shape of the Majorana end states is subject to strong renormalization
and can deviate strongly from the noninteracting 
limit.\cite{gangadharaiah:2011,stoudenmire:2011,sela:2011,lutchyn:2011}
Here we add a further element of caution: We show that the assumed equivalence
of HLL and SLL breaks down for any interacting electron system, and that 
major differences arise in the response functions. These differences persist
if the systems are brought in the proximity to a superconductor, and may lead to 
a different response of the bulk as well as of the Majorana properties in both types
of systems.

Because of the specific magnetic ordering in the SLL, we find that its electronic excitations can be
divided into a \emph{regular sector} that exhibits a typical LL (or HLL) type behavior, as well as an 
\emph{irregular sector} that exhibits a more special type of physics. In the non-interacting limit the 
quasiparticles in the regular sector 
are typical chiral-propagating modes, while the irregular sector is gapped. There is still some propagation 
in this sector because
the left-moving spin-up ($L\up$) and right-moving spin-down ($R\dw$) modes, characterized by the boson fields
$\phi_{L\up}$ and $\phi_{R\dw}$, are not separately gapped; 
instead, they are pinned together through
$\phi_{L\up} = \phi_{R\dw}$. 
In this sense, the gapped sector exhibits similarities to the LL edge states, for which
the reflection of the quasiparticles at the end of the wire pins the $L$ and $R$ movers 
together.\cite{eggert92,wang94,eggert95,fabrizio95,eggert96,eggert97,mattson97}
The difference between the two is that the pinning of $\phi_{L\up}$ and $\phi_{R\dw}$ for the irregular sector of the SLL 
is a bulk property, while for the edge of the LL the pinning of the $L$ and $R$ modes is local and happens only at the 
end of the wire. Thus, the modification of power-law exponents in the SLL irregular sector is similar to the modification 
of the end-tunneling exponents for a LL.
In contrast, this physics is absent in a HLL.

In this paper, we analyze the spectral properties of a SLL in comparison with those of the regular LLs and 
the HLL. In particular, we analyze the dependence of the two-point
retarded Green's function on position and frequency, 
as well as the dependence of the DOS on frequency.

Our first result is a closed and simple analytical form for the frequency dependence
of the local retarded Green's functions for the spinless LL, the HLL, as well as for the
regular sector of the SLL.
While the analytical expressions of the Green's function of a LL in position-time $(x,t)$ space
as well as momentum-frequency $(k,\omega)$ space are well known,\cite{gogolin,giamarchi,meden:1992,voit:1993,kivelson:2003}
the position-frequency $(x,\omega)$ dependence has been either calculated directly numerically 
by Fourier transform\cite{eggert06} or has been obtained by an efficient recursion method.\cite{se}
However, to our knowledge, an explicit analytic expression has not been provided yet.
Here we succeed in directly evaluating the Fourier 
transform of the position-time dependent Green's function, and we obtain a closed compact 
form for the position-frequency dependence of the imaginary time and retarded Green's functions.
For the spinful LL as well as for the irregular sector of the SLL we cannot provide such a closed
form. Instead, we give the results in form of simple convolution integrals.

Second, we use the obtained results to derive the frequency dependent local (or tunneling) DOS. 
We compare the results for the SLL to that of the standard LL 
and of the HLL. 
For HLLs and regular LLs the irregular contributions are absent. The DOS does not show a gap and follows 
the usual power-law dependence.
The DOS of the SLL is given by the sum of the contributions for the regular sector and the gapped irregular sector
and therefore exhibits a pseudogap. Inside the pseudogap the DOS shows a power-law dependence, characteristic 
of the regular sector. 
However, at the gap edge, the DOS is dominated by the irregular sector and depends on the value 
of the interaction parameter, 
such that for weak interactions it exhibits a typical divergence, while for strong interactions this divergence 
goes to zero.
Consequently, we propose to use the dependence of the DOS on frequency to test the apparition
of a magnetically ordered phase due either to the hyperfine interaction with the lattice of nuclear spins or to the 
spin-orbit interactions. The measurement of the DOS can provide a direct access to the size of the pseudogap, 
while the form of the gap-edge singularity may allow to infer the strength of the interactions. 
A similar behavior of the DOS can be found in 1D systems with charge density wave 
order.\cite{schuricht:2008,schuricht:2011a,schuricht:2011b}

We note, finally, that our work has been followed-up by Ref. \onlinecite{schuricht:2011c}, which 
contains a computation of the spectral properties of the SLL in $(k,\omega)$ space,
and a derivation of the DOS near the pseudogap edge beyond the harmonic approximation 
used in the present work.

The plan of the paper is as follows: 
In Sec. \ref{sec:models}, we introduce the various models under consideration, 
corresponding to the standard LL, the HLL, and the SLL. 
Section \ref{sec:gf} constitutes the core of the paper. In this section we 
present our results for the two-point $(x,\omega)$ Green's functions. 
In Sec. \ref{sec:dos}, we compute the DOS of the various LLs.
We provide a summary of our results in Sec. \ref{sec:conclusions}. Three appendices are also included,
containing details of the Green's functions calculations (Appendix \ref{ap:retarded}), the DOS calculations
(Appendix \ref{ap:alpha}), as well as the derivation of an important identity (Appendix \ref{ap:rho_integral}).


\section{Models}
\label{sec:models}

We start by introducing the models that we will consider in our analysis. These models consist of different 
variations of the LL model. We summarize here only the main features of these models, and for a thorough 
discussion of the LL physics we refer the reader, for example, to Refs.~\onlinecite{gogolin} and \onlinecite{giamarchi}.
Our notations and normalizations for the bosonic fields follow the conventions of Ref.~\onlinecite{giamarchi}. 
We consider the following three types of LLs: the standard spinless and spinful LL; the helical LL appearing on the 
edge of topological insulators; and
the spiral LL obtained by the self-ordering transition due to the hyperfine coupling to nuclear spins,
or due to a commensurability transition in systems with spin-orbit interactions.


\subsection{Standard Luttinger liquids}

Interacting spinless electrons in a 1D conductor can be described
by the LL Hamiltonian
\be
\label{eq:hll}
	H_{LL}
	=
	\int \frac{dx}{2\pi} v \left[ \frac{1}{K}(\nabla\phi)^2 + K(\nabla\theta)^2 \right],
\ee
where $\phi$ and $\theta$ are bosonic fields such that $-\nabla\phi/\pi$ measures
the electron density fluctuations in the system and
$\nabla \theta/\pi$ is canonically conjugate to $\phi$. We set $\hbar = 1$ throughout the paper.
The $x$ integral is performed over the system length $\mathcal{L}$. We assume $\mathcal{L}$ to
be much longer than the width of the low-energy wave packets, $\mathcal{L} k_F \ll 1$,
where $k_F$ is the Fermi momentum, which allows us to use a continuum description.
The LL parameter $K$ incorporates the effects of the electron-electron interactions, $K=1$
corresponds to non-interacting electrons, and $0<K<1$ corresponds to repulsive
electronic interactions. The quantity $v$ is the velocity of the bosonic model, which for 
an ideal LL is given by $v = v_F / K$, with $v_F$ being the Fermi velocity.

In this theory, the electron operator $\psi$ is written as $\psi = \psi_L + \psi_R$,
where $\psi_L$ corresponds to the left-moving modes with momenta close to $-k_F$,
and $\psi_R$ to the right-moving modes with momenta close to $k_F$.
These fermion operators are related to the boson fields as 
\begin{equation} \label{eq:defpsi}
	\psi_{r}(x) = \frac{\eta_r}{\sqrt{2\pi a}} \e^{i r k_F x} \e^{-i r \phi_r(x)},
\end{equation}
with $r = L,R = -,+$, and with the chiral boson fields
\begin{equation} \label{eq:defphi}
	\phi_r(x) = \phi(x) - r \theta(x).
\end{equation}
The quantity $a$ in Eq. \eqref{eq:defpsi} is a short distance cutoff, limited from below by the lattice constant,
and $\eta_r$ is the ladder operator, or Klein factor, whose effect is to lower the
$r$ particle number by 1.

For electrons with spin, the Hamiltonian of the 1D conductor in the LL regime
takes the form $H = H_c + H_s$, where $H_{c,s}$ have the same form as
$H_{LL}$ in Eq. \eqref{eq:hll},
\be \label{eq:hll_spin}
	H_{\nu}
	=
	\int \frac{dx}{2\pi} v_\nu \left[ \frac{1}{K_\nu}(\nabla\phi_\nu)^2 + K_\nu(\nabla\theta_\nu)^2 \right],
\ee
for $\nu = c, s$ labeling the charge and spin degrees of freedom, and $v_{\nu}=v_F/K_{\nu}$. The decoupling of $H_c$ and $H_s$
is the consequence of the spin-charge separation in 1D.
Here $K_{c,s}=1$ corresponds to the noninteracting case, $0<K_c<1$ to repulsive electron interactions,
and $K_s \neq 1$ to a broken spin SU(2) symmetry. The presented LL theory applies
only for $K_s \ge 1$, as a value of $K_s$ smaller than 1 would lead to the opening of a spin-gap and so to different physics.
The normalizations are chosen such that $-\nabla \phi_{c,s} \sqrt{2}/\pi$ give the magnitude of the charge- and 
spin-density fluctuations. The electron operators can again be separated into left- and right-movers,
$\psi_{\sigma} = \psi_{L\sigma} + \psi_{R\sigma}$, where
\begin{equation}  \label{eq:psi_spin}
	\psi_{r \sigma}(x) = \frac{\eta_{r\sigma}}{\sqrt{2\pi a}} \e^{i r k_F x} \e^{-i r \phi_{r\sigma}(x)},
\end{equation}
with the additional spin index $\sigma = \up,\dw = +,-$ and the
boson fields
\begin{equation} \label{eq:phi_spin}
	\phi_{r\sigma}(x) = \frac{1}{\sqrt{2}}
	\bigl[\phi_c(x) - r \theta_c(x) + \sigma \bigl(\phi_s(x) - r \theta_s(x) \bigr) \bigr].
\end{equation}
This expansion of $\phi_{r\sigma}(x)$  contains both charge and spin fields, indicating that an electron
is a superposition of both spin and charge degrees of freedom.


\subsection{Helical Luttinger liquid}
\label{sec:hll_model}

The quantum spin-Hall effect appears in various systems with strong spin-orbit interaction and preserved 
time-reversal symmetry. While the band structure has been long known,\cite{dyakonov81,dyakonov82,pankratov87}
such systems have found much renewed interest recently because it was recognized that the electron state 
indeed describes the quantum spin-Hall effect, a topological state of matter characterized by a bulk gap and gapless edges.\cite{qshe,hasan:2010}
Recent experiments on HgTe quantum wells have 
provided direct evidence for nonlocal transport
in the quantum spin-Hall regime in the absence of any external magnetic field, in agreement with the theory.\cite{molenkamp1,molenkamp2}
Due to the strong spin-orbit coupling, the left- and the right-movers of the gapless edge modes 
carry opposite spin. These edge chiral excitations 
are described by a separate class of one-dimensional LLs, the HLLs.\cite{wu06,xu06} 
Contrary to the chiral LLs at the edges
of fractional quantum Hall systems,
these edges states do not break time-reversal symmetry.

The linearized Hamiltonian of the non-interacting HLL in the fermionic language can be written as
\be\label{eq:hhll}
H_{HLL}^0=-v_F\int dx\;(\psi_{R\up}^\dag i\p_x\psi_{R\up}-\psi_{L\dw}^\dag i\p_x\psi_{L\dw}).
\ee
One can easily describe the interactions between fermions of the same species,
as well as the forward scattering processes in the bosonized language.\cite{wu06,hou:2009,strom:2009,teo:2009,tanaka:2009,liu:2011} 
Using Eq. (\ref{eq:psi_spin}) the Hamiltonian in Eq. (\ref{eq:hhll}) can be written
in a bosonized form.
Moreover, in a HLL, the spin and chirality indices coincide due to the strong spin-orbit coupling, and, as one can 
see directly in Eq.~(\ref{eq:hhll}), the spin index is redundant.
It is therefore more convenient to treat the Hamiltonian of the HLL as an effectively spinless LL by introducing the fields
$\Phi=(\phi_{R\up}+\phi_{L\dw})/\sqrt{2}$ and $\Theta=(\phi_{L\dw}-\phi_{R\up})/\sqrt{2}$.
When including the interactions, the bosonized form of the Hamiltonian of the
interacting HLL becomes\cite{wu06,hou:2009,strom:2009,teo:2009,tanaka:2009,liu:2011}
\be\label{eq:hhll_int}
	H_{HLL}=\int \frac{dx}{2\pi} v\left[\frac{1}{K_{HLL}}(\nabla\Phi)^2+K_{HLL}(\nabla\Theta)^2\right],
\ee
where we have introduced the parameter $K_{HLL}$, which is a generalized LL parameter taking into account the 
interactions, and $v=v_F/K_{HLL}$, the velocity of the collective excitations.
Recent Quantum Monte Carlo simulations have shown that this HLL description of the helical conductor remains 
indeed valid as long as the interaction strength remains smaller than the band gap 
protecting the edge states from the higher-dimensional bulk states.\cite{hohenadler:2011a,hohenadler:2011b}


\subsection{Spiral Luttinger liquid}

The presence of nuclear spins in a 1D conductor can substantially modify the
LL behavior. It was shown in Refs. \onlinecite{bsl1} and \onlinecite{bsl2} that the hyperfine
interaction between the nuclear and electron spins can indeed trigger a strong feedback reaction
between nuclear spins and the electron modes, leading to ordered phases in both
subsystems. This order is manifested in the apparition of a nuclear helimagnet that is stabilized by its coupling to
the electron system. In turn, the resulting spiral nuclear magnetic field generates a
relevant sine-Gordon perturbation for the electrons. The Hamiltonian
that captures this electron physics can be written as\cite{bsl1,bsl2}
\begin{align}
	H_{el} = \int \frac{dx}{2\pi} 
	\biggl\{ 
		&\sum_{\kappa=\pm}
		v_\kappa \Bigl[(\nabla \phi_\kappa)^2 + (\nabla \theta_\kappa)^2 \Bigr]
\nonumber \\
		&+ 
		\frac{B_{\rm eff}}{a}\cos\left(\sqrt{2K}\phi_+\right)\biggr\},
\label{eq:H_pm}
\end{align}
with
\begin{eqnarray}
	v_+ &&= (v_cK_c+v_sK_s^{-1})/K\\
	v_- &&= (v_cK_s^{-1}+v_sK_c)/K,
\label{eq:v-}
\end{eqnarray}
where we have introduced
\begin{equation}
K = K_c + K_s^{-1}.
\end{equation}
In Eq. \eqref{eq:H_pm} we have neglected unimportant marginal terms coupling the ``$+$'' and ``$-$'' fields.

The field $B_{\rm eff}$ in Eq. (\ref{eq:H_pm}) represents the effective Overhauser magnetic field induced by the nuclear spiral order.
The fields
 $\phi_\pm$ and $\theta_\pm$ are related to the charge and spin boson fields by\cite{bsl1,bsl2,typo}
\begin{align}
\label{eq:phi_c}	\phi_c
	&= \frac{1}{\sqrt{K}} \left[ K_c \phi_+ - \sqrt{\frac{K_c}{K_s}} \phi_- \right],
\\
	\theta_s
	&= \frac{1}{\sqrt{K}} \left[ \frac{1}{K_s} \phi_+ + \sqrt{\frac{K_c}{K_s}} \phi_- \right],
\\
\label{eq:theta_c}
	\theta_c
	&= \frac{1}{\sqrt{K}} \left[ \theta_+ - \frac{1}{\sqrt{K_c K_s}} \theta_- \right],
\\
\label{eq:phi_s}
	\phi_s
	&= \frac{1}{\sqrt{K}} \left[ \theta_+ + \sqrt{K_c K_s} \theta_- \right].
\end{align}
An identical sine-Gordon renormalization was also identified in low-density 1D electron conductors with
spin-orbit interaction, such as GaAs or InAs wires,\cite{bjkl} 
and can also appear in carbon nanotubes.\cite{ksbl1,ksbl2}
Promising candidates are furthermore Ge/Si core/shell nanowires as they combine
an unusually large, controllable spin-orbit interaction,\cite{kloeffel:2011,hao:2010}
a good hole carrier confinement in the core, and high mobilities.\cite{lu:2005,xiang:2006,park:2010,lee:2010}

The Overhauser field is then replaced in the semiconductor wires by a uniform magnetic field,
and in the carbon nanotubes by the combination of intrinsic spin-orbit splitting together with an electric field.
The relevant renormalization occurs as a commensurability transition when the
electron density is tuned such that $k_F = k_{so}$, with $k_{so}$ being the inverse spin-orbit length.

The cosine term in Eq.~(\ref{eq:H_pm})  is strongly relevant and the field $\phi_+$
becomes pinned to a constant. Excitations in this sector are characterized by a gap $\Delta$.
The nuclear Overhauser field binds indeed one half of the electron
modes into a spin and charge mixing electron density wave with a spiral spin polarization that
follows the nuclear spin helix. Through the hyperfine interaction the stability of the nuclear helix
is further enhanced by the SLL.
The combined electron--nuclear-spin order was predicted to be stable up to
temperatures of 10--100 mK for experimentally available
GaAs quantum wires\cite{pfeiffer,yacoby02,yacoby05,yacoby08} and $^{13}$C substituted
single-wall carbon nanotubes.\cite{simon,ruemmeli,churchill_1,churchill_2}

The  remaining half of the
electron modes can be described by a modified LL, characterized by a different LL parameter.
We have termed the resulting low-energy Hamiltonian
the ``spiral Luttinger Liquid'' (SLL).
Similarly to the HLL discussed in Sec. \ref{sec:hll_model}, the SLL possesses
spin-filtered left- and right-movers, yet the SLL exhibits
important differences with respect to the HLL.
We will discuss these differences in detail in the following.

Let us mention that Hamiltonian \eqref{eq:H_pm} does not directly apply to the 
case carbon nanotubes, as the latter require a two-band description that captures the 
physics of the two Dirac valleys of the nanotubes.\cite{egger:1997,kane:1997}
However, as was shown in Ref. \onlinecite{bsl2}, cosine terms such as in Eq. \eqref{eq:H_pm}
open gaps in each Dirac valley separately and overrule the otherwise strong inter-valley 
coupling caused by the electron-electron interaction. The SLL physics of carbon nanotubes
can be, therefore, captured by two independent copies of Eq. \eqref{eq:H_pm}, one for 
each valley. In the following, therefore, we restrict entirely to the single-band
model of Eq. \eqref{eq:H_pm} and note that for most of the measurable response functions
the difference to nanotubes amounts in a mere factor 2 in the results.


\subsection{Differences between Spiral and Helical Luttinger liquids}

The main difference between the SLL and the HLL is the origin of the spin-filtered 
conduction modes. While for the HLL they are a consequence of the special 2D or 3D band 
structure, for the SLL they result from the 
opening of a spin-dependent pseudogap $\Delta$ in an otherwise conventional spinful 1D bulk 
conductor. In the non-interacting limit the separation between the gapless, 
spin-filtered modes $(L\dw)$, $(R\up)$ and the gapped modes $(L\up)$, $(R\dw)$ is 
perfect, and for energies $|\omega| \ll \Delta$ the physics of the SLL, governed by 
$(L\dw)$, $(R\up)$, is identical to the HLL.
For interacting electrons, however, parts of the gapless phase, described 
by the boson modes $\phi_-$ and $\theta_-$ affect the $(L\up)$, $(R\dw)$ sector as well
and lead to what we call the ``irregular'' response even at low energies.

To see this in detail, let us project the initial electron Hamiltonian onto the gapless 
(``$-$'') sector and rewrite the Hamiltonian \eqref{eq:H_pm} in the standard form of a 
spinless LL \eqref{eq:hll}, reminiscent of the HLL. We start with rescaling
the fields $\phi_-,\theta_-$ as
\be
	\phi   = -\sqrt{\frac{2K_c}{KK_s}} \phi_-, \quad
	\theta = -\sqrt{\frac{KK_s}{2K_c}}\theta_-,
\ee
which allows us to rewrite the gapless sector of Eq. \eqref{eq:H_pm} as
\begin{equation} \label{eq:H_SLL}
	H_{SLL} = \int \frac{dx}{2\pi} v_{SLL} \left[ \frac{1}{K_{SLL}} (\nabla\phi)^2 + K_{SLL} (\nabla\theta)^2 \right],
\end{equation}
with the parameters
\begin{equation} \label{eq:Kv_SLL}
	K_{SLL} = 2K_c/(K_cK_s+1),
	\quad
	v_{SLL} = v_-.
\end{equation}
If we neglect the ``$+$'' fields for the time being, the boson fields are projected onto the gapless
sector as
\begin{align}
	\phi_c
	&= - \theta_s \to \frac{1}{\sqrt{2}} \phi,
\\
	\theta_c
	&\to \frac{\sqrt{2}}{K K_s} \theta,
\\
	\phi_s
	&\to -\frac{\sqrt{2} K_c}{K} \theta,
\end{align}
and the left- and right-movers become
\begin{align}
	\phi_{L\dw} &\to \phi + \theta,
\label{eq:phi_Ldw}
\\
	\phi_{R\up} &\to \phi - \theta,
\label{eq:phi_Rup}
\\
	\phi_{L\up} &= - \phi_{R\dw} \to -\frac{\bar{K}}{K} \theta,
\label{eq:phi_Rdw_phi_Lup}
\end{align}
where $\bar{K} = K_c - K_s^{-1}$.

While the form of $\phi_{L\dw}$ and $\phi_{R\up}$ is similar to that in Eq. \eqref{eq:defphi}, the
presence of the propagating $\theta$ field in the expression of $\phi_{R\dw}$ and $\phi_{L\up}$ is, at first 
glance, surprising, since the exponentials of these boson fields correspond to the nonpropagating fermion 
operators of the pinned spiral density wave. However, such a single-electron
picture is inappropriate for an interacting system. Instead, for an interacting 
system, a left- moving electron corresponds to a superposition
of left-moving particle-like and right-moving hole-like bosonic modes that together carry unit charge and spin.
Only in the noninteracting limit does the amplitude of the right-moving excitations vanish. 
Moreover, the gap $\Delta$ builds up from the elementary bosonic modes, which carry only a fraction of the electron charge.
The connection between $\phi_{L\up}$ and $\phi_{R\up}$ is therefore a natural consequence of electronic interactions. 
A similar connection holds between
$\phi_{R\dw}$ and $\phi_{L\dw}$. Since $\phi_{R\dw} = - \phi_{L\up}$ (after projection),
only the antisymmetric part $\theta$ between the gapless $\phi_{R\up}$ and $\phi_{L\dw}$
can be excited, which is encoded in Eq. \eqref{eq:phi_Rdw_phi_Lup}. 

This peculiar $\theta$ dependence is the major difference between the
SLL and the HLL, and leads to the irregular contributions to the 
response functions that are detailed below. It is a pure interaction effect and, indeed, the prefactor 
$\bar{K}$ of the irregular $\theta$ correlators of the fields in Eq. \eqref{eq:phi_Rdw_phi_Lup}
vanishes in the non-interacting limit.


\section{Green's functions in $(x,\omega)$ space}
\label{sec:gf}

This section describes the derivation of one of the main results of the paper, the form of the two-point
Green's functions in $(x,\omega)$ space, which constitutes the basis for the description of the spectral 
single-particle properties of the conductors. 
While much is known about the Green's functions in
position--time $(x,t)$ (and imaginary time $\tau = it$) space,\cite{gogolin,giamarchi}
as well as in momentum--frequency $(k,\omega)$ space,\cite{meden:1992,voit:1993} to our knowledge
the correlators in the $(x,\omega)$ space have been to date determined only numerically. \cite{eggert06,se}
Here we provide analytical, closed expressions for the spinless LL, the HLL, and the regular sector of the SLL 
Green's functions.
For the spinful LL as well as for the irregular sector of the SLL
we provide integral formulas in form of convolution integrals.
We focus here on imaginary-time ordered Green's functions and on
real-time retarded Green's functions, yet our calculations can be easily repeated for all
other types of Green's function.

We define in the standard way the imaginary time (Matsubara) Green's function in $(x,\tau=it)$ space for a 
translationally invariant stationary system as
\begin{equation}
	\tilde{G}_{\bar{r},\bar{r}'}(x,\tau) = - \mean{T_\tau \psi_{\bar{r}}^\dagger(x,\tau) \psi_{\bar{r}'}(0,0)},
\end{equation}
where $\bar{r} = (r,\sigma,\dots)$ describes the chirality, spin, and any other quantum number characterizing 
the electron operators, and $T_\tau$ is the usual time order operator.
The real time Green's functions are expressed using the elementary greater and lesser functions
\begin{align}
	G^{>}_{\bar{r},\bar{r}'}(x,t) &= - i\mean{\psi_{\bar{r}}^\dagger(x,t) \psi_{\bar{r}'}(0,0)},
\\
	G^{<}_{\bar{r},\bar{r}'}(x,t) &= + i\mean{\psi_{\bar{r}'}(0,0) \psi_{\bar{r}}^\dagger(x,t)},
\end{align}
which allow us, for instance, to express the retarded Green's function as
\begin{equation}
	G^{\rm ret}_{\bar{r},\bar{r}'}(x,t) =  \vartheta(t)
	[ G^{>}_{\bar{r},\bar{r}'}(x,t) - G^{<}_{\bar{r},\bar{r}'}(x,t) ],
\end{equation}
with $\vartheta(t)$ the unit step function.
The Fourier transforms to frequency space are given by
\begin{align}
	\tilde{G}_{\bar{r},\bar{r}'}(x,i \omega_n)
	&= \int_{-\infty}^{\infty} d\tau \, \e^{i \omega_n \tau} \, \tilde{G}_{\bar{r},\bar{r}'}(x,\tau),
\label{eq:FT_Matsubara}
\\
	G_{\bar{r},\bar{r}'}^{\alpha}(x,\omega)
	&= \int_{-\infty}^\infty dt \, \e^{i \omega t} \, G_{\bar{r},\bar{r}'}^{\alpha}(x,t),
\label{eq:FT_real_time}
\end{align}
where $\alpha \in \{>,<,\rm ret\}$. We restrict our analysis to zero temperature.


\subsection{Standard Luttinger liquids}
\label{sec:GF}


\subsubsection{Spinless Luttinger liquid}

For the standard LL,
the calculation of imaginary-time fermionic correlators reduces to
calculating the correlators between the boson fields $\phi$ and $\theta$.
To this end, it is convenient to introduce left- and right-moving eigenmodes
\begin{equation}\label{eq:tilde_phi}
	\tilde{\phi}_r = \frac{1}{\sqrt{2K}} (\phi - r K \theta),
\end{equation}
which commute with each other. The evaluation of their correlators in $(x,\tau)$ space
is well described in
textbooks.\cite{gogolin,giamarchi} At zero temperature, for a translationally invariant system described 
by the Hamiltonian in Eq.~\eqref{eq:hll}, we have
\begin{equation} \label{eq:log_correlator}
	\mean{[\tilde{\phi}_r(x,\tau) - \tilde{\phi}_r(0,0)]^2}
	= -\ln\left[\frac{x - i r v \tau}{i r a} \right].
\end{equation}
This result is valid in the limit $|x|,v|\tau| \gg a$, in which bosonization applies.
The electron correlators are then evaluated using the relation
\begin{align}
	&\mean{\e^{i [\lambda \phi(x,\tau) + \mu \theta(x,\tau)]} \e^{- i [\lambda \phi(0,0) + \mu \theta(0,0)]}}
\nonumber\\
	&= \e^{\frac{1}{2} \mean{[\lambda\phi(x,\tau)+\mu\theta(x,\tau) - \lambda\phi(0,0) - \mu \theta(0,0)]^2}}
\label{eq:bosonic_correlators}
\end{align}
for arbitrary constants $\lambda$ and $\mu$.
Using Eqs.~\eqref{eq:defpsi} and \eqref{eq:defphi}, for the
spinless LL  this leads to
\begin{align} \label{eq:G_tau0}
	\tilde{G}_{r,r}(x,\tau)
	= - \frac{\e^{i r k_F x}}{2\pi a}
	\left[\frac{-i a}{x - i v \tau}\right]^{\gamma_r}
	\left[\frac{i a}{x + i v \tau}\right]^{\gamma'_r},
\end{align}
with
\begin{equation}
	\gamma_r = \gamma_{-r}' = \frac{K+K^{-1}-2 r}{4}.
\end{equation}
Note that $|\gamma_r' - \gamma_r| = 1$.
Cross-correlators $\tilde{G}_{r,r'}$ with $r \neq r'$ vanish as they do not preserve the
number of left- and right-movers.

The Fourier transformation to $\tilde{G}_{r,r}(x,i\omega_n)$ can directly be evaluated using standard
integral tables [Ref. \onlinecite{gradshteyn}, 3.384.9], and leads to a
Whittaker function. Yet since the properties of the Whittaker function might be unfamiliar
to the reader, we propose here a modified calculation that expresses the result in terms
of the more commonly known Bessel functions, in a combination that is equivalent to the Whittaker function.

Since this calculation yields one of the main results of this paper, we provide here the details
for the imaginary time Green's function calculation. The analogous calculation for the retarded Green's function 
can be found in Appendix \ref{ap:retarded}. As explained below, the correspondence between these two Green's functions, 
usually given by the Wick rotation, is actually problematic here (and would be 
even more obscure if we would use the Whittaker function), and an independent calculation for the two Green's 
functions is necessary.

We start by rewriting the imaginary-time Green's function as
\begin{equation} \label{eq:G_tau}
	\tilde{G}_{r,r}(x,\tau)
	=
	i \frac{\e^{i r k_F x}}{4\pi \gamma}
	\bigl(\partial_{rx} + \partial_{i v \tau} \bigr)
	\left[\frac{a}{rx-i v \tau} \frac{a}{rx+i v \tau}\right]^\gamma,
\end{equation}
with $\gamma = \gamma_R = (K+K^{-1}-2)/4$.
We use differentiation to reduce the
exponent of the singularity in the integrand, which provides us with an elegant
way to regularize the unphysical ultraviolet divergences.
Indeed, we have $0<\gamma<1$ for $K>0.17$,
which will assure convergence of the integrals below. For smaller values of $K$,
we can use an analytic continuation of the result in $\gamma$ and neglect any occurring
divergence, or we can use again the regularization by
further applications of the differentiation.
We thus need to calculate the integral
\begin{equation} \label{eq:I}
	I(x,i\omega_n) = \int_{-\infty}^\infty
	d\tau \, \e^{i \omega_n \tau} \left[\frac{a}{x^2+ (v \tau)^2} \right]^\gamma,
\end{equation}
which can be done using standard integral tables [Ref. \onlinecite{gradshteyn}, 8.432.5].
The result is given by
\begin{equation} \label{eq:I_sol}
	I(x,i\omega_n) = \frac{2a\sqrt{\pi}}{v\Gamma(\gamma)}
	\left(\frac{2|x| v}{|\omega_n|a^2}\right)^{\frac{1}{2}-\gamma} K_{\gamma-\frac{1}{2}}(|x \omega_n|/v),
\end{equation}
where $K_\alpha(z)$ is the modified Bessel function and $\Gamma(z)$ Euler's Gamma function.
To obtain the Green's function we need to evaluate also the derivatives in Eq.~\eqref{eq:G_tau}.
The time derivative yields a factor $-\omega_n/v$, while the derivative with respect to $r x$
is evaluated using the identity
\begin{equation} \label{eq:K_id}
	\frac{d}{dz} z^{-\alpha} K_\alpha(z)
	= z^{-\alpha}K_{\alpha+1}(z),
\end{equation}
from which we obtain the result
\begin{align} \label{eq:G_Matsubara_LL}
	&\tilde{G}_{r,r}(x,i\omega_n)
	= \frac{-\e^{i r k_F x} }{2\sqrt{\pi} \Gamma(\gamma+1)} \frac{i \omega_n a}{v^2}
	\left(\frac{2|x|v}{|\omega_n|a^2}\right)^{\frac{1}{2}-\gamma}
\nonumber \\
	&\times
	\left[ K_{\gamma-\frac{1}{2}}(|x \omega_n|/v) - \sign(r x \omega_n) K_{\gamma+\frac{1}{2}}(|x\omega_n|/v) \right].
\end{align}
This expression of the Green's function holds for a real $\omega_n$.
The analytic continuation to the retarded Green's function by the Wick rotation is, however, not well defined, 
due to the nonanalytic dependence on $|\omega_n|$, which arises because taking the zero temperature limit and the analytic 
continuation do not commute.
We circumvent this problem by directly computing $G^{ret}(x,\omega)$ using contour integrations.
The details are summarized in Appendix \ref{ap:retarded}. The result is
\begin{align} \label{eq:G_ret_LL}
	&G^{ret}_{r,r}(x,\omega)
	= \frac{-\e^{i r k_F x} }{2\sqrt{\pi} \Gamma(\gamma+1)} \frac{\omega_+a}{v^2}
	\left(\frac{2i|x|v}{\omega_+a^2}\right)^{\frac{1}{2}-\gamma}
\nonumber \\
	&\times
	\left[ K_{\gamma-\frac{1}{2}}(|x| \omega_+/iv) - \sign(r x) K_{\gamma+\frac{1}{2}}(|x|\omega_+/iv) \right],
\end{align}
with $\omega_+ = \omega + i 0$. As a function of $\omega$, the Bessel functions have branch
cuts along the negative imaginary axis, i.e., at the Matsubara frequencies $\omega_n < 0$.
This shows that indeed the analytic continuation between the real and imaginary time Green's functions
via the Wick rotation cannot be naively used, justifying the independent calculation.


\subsubsection{Luttinger liquid with spin}
\label{sec:Greens_LL}

The LL model for true, spinful electrons can be obtained by considering two copies of the LL model (for the 
charge and spin fields), as described by the Hamiltonians in Eq. \eqref{eq:hll_spin}. The
electron operators are given by Eq.~\eqref{eq:psi_spin}, together with Eq.~\eqref{eq:phi_spin}.
Since the charge boson fields commute with the spin boson fields, the evaluation of the
bosonic correlators as expressed in Eq.~\eqref{eq:bosonic_correlators} factorizes into two exponentials,
each of them depending only on one of the charge and spin fields. Thus Eq. \eqref{eq:G_tau0} can be replaced by
\begin{align}
	\tilde{G}_{r\sigma,r\sigma}(x,\tau)
	= - &\frac{\e^{i r k_F x}}{2\pi a}
	\left[\frac{-i a}{x - i v_c \tau}\right]^{\gamma_{r,c}}
	\left[\frac{i a}{x + i v_c \tau}\right]^{\gamma'_{r,c}}
\nonumber \\
	&\times
	\left[\frac{-i a}{x - i v_s \tau}\right]^{\gamma_{r,s}}
	\left[\frac{i a}{x + i v_s \tau}\right]^{\gamma'_{r,s}},
\label{eq:G_spin}
\end{align}
with
\begin{equation}
	\gamma_{r,\nu} = \gamma_{-r,\nu}' = \frac{K_\nu +K_\nu^{-1}-2 r}{8},
\end{equation}
where $\nu = c,s$.
As before, the cross-correlators with $(r,\sigma) \neq (r',\sigma')$ vanish,
and $|\gamma_{r,\nu} - \gamma_{r,\nu}'|=1/2$.
Generally we have $v_c \neq v_s$, and the Fourier transform cannot be calculated
using Eq. \eqref{eq:I}. To our knowledge, a general solution to this Fourier transform
has not yet been found. However, it can be evaluated exactly in the special cases of
$v_c = v_s$ and of $x=0$. If the velocities coincide, the Green's functions can be obtained by the same 
type of calculation as the Green's functions for a spinless LL, and are given by Eqs. \eqref{eq:G_Matsubara_LL} 
and \eqref{eq:G_ret_LL}, on the replacements $r \to (r,\sigma)$ and
$\gamma \to \gamma_{R,c}+\gamma_{R,s} = [(K_c+K_s)+(K_c^{-1}+K_s^{-1})-4 r] / 8$.
The limit $x=0$, important for the calculation of the DOS in Sec.~\ref{sec:dos}, can be obtained 
in a similar manner, noting that one needs to evaluate an integral of the same type as the one in Eq. \eqref{eq:I}, 
on the additional replacement $v \to \sqrt{v_c v_s}$, and on taking the limit $x \to 0$.

For general velocities and for $x\neq 0$, we note that Eq. \eqref{eq:G_spin} can be viewed
as the product of two spinless Green's functions with labels $c$ and $s$. Hence its
Fourier transform can be written as the convolution
\begin{align}
	&\tilde{G}_{r\sigma, r\sigma}(x,i\omega_n)
	= \frac{-\e^{i r k_F x} (2|x|/a)^{1-\gamma_c-\gamma_s}}{4\pi v_c v_s\Gamma(\gamma_c+1)\Gamma(\gamma_s+1)}
\nonumber \\
	&\times
	\int d\omega_n' f_{r,c}(x,i\omega_n') f_{r,s}(x,i\omega_n-i\omega_n'),
\label{eq:G_conv}
\end{align}
with
\begin{align}
	&f_{r,\nu}(x,i\omega_n)
	= \frac{\omega_n a}{v_\nu} \left(\frac{|\omega_n| a}{v_\nu}\right)^{\gamma_\nu -\frac{1}{2}}
\nonumber\\
	&\times
	\left[ K_{\gamma_\nu-\frac{1}{2}}(|x \omega_n|/v_\nu) - \sign(r x \omega_n) K_{\gamma_\nu+\frac{1}{2}}(|x\omega_n|/v_\nu) \right].
\label{eq:f_conv}
\end{align}
While Eqs. \eqref{eq:G_conv} and \eqref{eq:f_conv} do not provide direct insight into the analytic shape of the
Green's functions of the spinful LL, due to the
exponential convergence of the integrals they are convenient for numerical evaluation.
A similar expression holds for $G^{ret}_{r\sigma,r\sigma}(x,\omega)$, yet with a
slower convergence because of the power-law decay of the Bessel function along the imaginary
axis.

\subsection{Helical Luttinger liquid}

The helical edge states arise from the special band structure of a topological insulator.
They are described by two counterpropagating edge modes with opposite spin (Kramers doublets), say
$(L\dw)$ and $(R\up)$. 
In the non-interacting system, an electron with spin $\up$ tunneling into the system has to move 
to the right and cannot enter into a coherent superposition of $(R\up)$ and the (nonexisting) $(L\up)$.
In the interacting system, however, chiral electrons are no longer the good eigenstates and the 
tunneling electron excites a coherent superposition of 
particle-like excitations of $(L\up)$ and hole-like excitations of $(R\dw)$, together carrying
unit charge and unit spin $\up$. 
The resulting theory is a fully
regular LL as described by the Hamiltonian in Eq.~\eqref{eq:hhll_int}. The Green's functions are
consequently given by Eqs. \eqref{eq:G_Matsubara_LL} and \eqref{eq:G_ret_LL}
with the replacement $\gamma \to \gamma_{HLL} = (K_{HLL} + K_{HLL}^{-1}-2)/4$.


\subsection{Spiral Luttinger liquid}

The SLL corresponds to the low-energy theory
of conduction electrons in the presence of the pseudogap $\Delta$.
Contrary to the HLL, the number of degrees of freedom is doubled, with one gapless 
sector and one gapped sector. 
If the SLL is caused by the nuclear spin ordering, there are two intrinsic energy scales in the system, the nuclear spin ordering temperature $T^*$
and the pseudogap $\Delta$ of the electrons, fulfilling for most of the semiconductor quantum wires $k_B T^* < \Delta$,
with $k_B$ the Boltzmann constant. We will assume henceforth that 
the temperature $T$ satisfies $T < T^*$ such that the SLL state remains stable.
We will also consider only frequencies $|\omega| \gg k_B T$, at which 
our zero-temperature analysis remains valid. 
For the SLL in spin-orbit interaction split semiconductor quantum wires, only the condition 
$|\omega| \gg k_B T$ is imposed as the scale $T^*$ is absent.
The distribution of energy scales allows us thus in particular
to explore the regime $|\omega| \sim \Delta$, which will be of much interest below.

In order to calculate the Green's function $\tilde{G}_{r\sigma,r\sigma}(x,\tau)$ in imaginary time, we first use 
Eqs. (\ref{eq:psi_spin}) and (\ref{eq:phi_spin}) to write the electron operators in terms of the charge and 
spin bosonic fields, and then use the change of basis given by
the set of Eqs. (\ref{eq:phi_c})--(\ref{eq:phi_s}) to write them in terms of $\phi_-,\theta_-,\phi_+,\theta_+$.
After these manipulations, the Green's function takes the form
\begin{equation}\label{eq:gr}
	\tilde{G}_{r\sigma,r\sigma}(x,\tau)=-\frac{\e^{irk_Fx}}{2\pi a} F_{r\s}^-(x,\tau)F_{r\s}^+(x,\tau),
\end{equation}
where
\begin{align}\label{eq:f-}
	&F_{r\s}^-(x,\tau) = 
\nonumber 
\\
	&
	\exp\Biggl(
		\frac{1}{4K} \Biggl\langle
			\Biggl[
				-(1+\sigma r)\sqrt{\frac{K_c}{K_s}}
				\bigl(\phi_-(x,\tau)-\phi_-(0,0)\bigr)
\nonumber 
\\
	&
				+\left(\frac{r}{\sqrt{K_cK_s}}+\sigma\sqrt{K_cK_s}\right)
				\bigl(\theta_-(x,\tau)-\theta_-(0,0) \bigr) 
			\Biggr]^2\,\Biggr\rangle
	\Biggr),
\end{align}
and
\begin{align}\label{eq:f+}
	&F_{r\s}^+(x,\tau)=
\nonumber
\\
	&\exp\Biggl(
		\frac{1}{4K}  \Biggl\langle
			\Biggl[
				\left(K_c-\frac{\sigma r}{K_s}\right)
				\bigl(\phi_+(x,\tau)-\phi_+(0,0)\bigr)
\nonumber
\\ 
	&
				+(\sigma-r)
				\bigl(\theta_+(x,\tau)-\theta_+(0,0)\bigr)
			\Biggr]^2
		\Biggr\rangle
	\Biggr).
\end{align}

\subsubsection{Regular sector}
\label{sec:regular_sector_GF}

Let us first consider the Green functions associated with the modes $(L\dw)$ and $(R\up)$, which corresponds to $r\sigma=1$. 
In this case, only the correlator $\langle(\phi_+(x,\tau)-\phi_+(0,0))^2\rangle$ enters in $F_{1}^+$. Since the field $\phi_+$ 
is gapped, $\langle(\phi_+(x,\tau)-\phi_+(0,0))^2\rangle$ decays exponentially
on the scale $\sqrt{x^2+v^2\tau^2} \sim v/\Delta$,
which implies that $F_1^+(x,\tau)\approx 1$. 
The correlation functions  between the  modes
$(L\dw)$ and $(R\up)$ are therefore described by a standard spinless LL theory. This corresponds to the
``regular sector'' of the SLL.
The Green's functions of the regular sector
are given by Eqs. \eqref{eq:G_Matsubara_LL} and
\eqref{eq:G_ret_LL}, on the replacements
$r = L \to (L\dw), r = R \to (R\up)$, and the new parameters
$v \to v_{SLL} = v_-$ as defined in Eq. \eqref{eq:v-} and
\begin{equation}
	\gamma \to \gamma_{reg}
	= \frac{K_{SLL} + K^{-1}_{SLL} - 2}{4},
\end{equation}
with $K_{SLL} = 2K_c/(K_c K_s+1)$.
Cross-correlators between the $(L\dw)$ and $(R\up)$ modes vanish.

\subsubsection{Irregular sector}

The correlation functions  between the modes
$(L\up)$ or $(R\dw)$ correspond to $r\sigma =-1$. The expression for these correlators is more complicated
since $F_{-1}^+$ in Eq. (\ref{eq:gr})
can no longer be approximated by unity due to the presence of the $\theta_+$ field in Eq. (\ref{eq:f+}),
which is conjugated to the pinned $\phi_+$ field and so strongly fluctuates.
We denote this as the ``irregular sector'' since the correlation functions cannot be inferred from a standard spinless LL analysis. 

The Green's function $\tilde{G}_{\bar{r},\bar{r}}(x,i \omega_n)$, with $\bar r\in \{\mbox{$(L\up)$}, \mbox{$(R\dw)$}\}$,
requires the calculation of the Fourier transform of the product $F_{-1}^-(x,\tau) F_{-1}^+(x,\tau)$,
for which we unfortunately we have not been able to find an analytical, closed form.
We can, however, evaluate $F^\pm_{-1}(x,i\omega_n)$ individually and express $\tilde{G}_{\bar{r},\bar{r}}(x,i \omega_n)$ 
in form of again a convolution integral similar to Eq. \eqref{eq:G_conv}. 
In addition, we will see in the next section that the knowledge of $F^\pm_{-1}(x=0,i\omega_n)$ 
is sufficient to obtain a closed form for the DOS. 
Therefore, we will focus here on the individual evaluations of $F^\pm_{-1}(x,i\omega_n)$.
While an analytic expression can be found for $F^-_{-1}(x,i\omega_n)$ for all $x$, this is possible for $F^+_{-1}(x,i\omega_n)$ 
only at $x=0$ (as shown in the next section). 
For $x\neq 0$ we provide the result in form of a convolution similar to Eq. \eqref{eq:G_conv}.

Let us start with $F^+_{-1}(x,i\omega_n)$.
In order to compute the asymptotic behavior of the associated Green functions, one needs to make a number of approximations. 
Following Ref. \onlinecite{starykh} we first replace the sine-Gordon term in Eq. (\ref{eq:H_pm}) by its quadratic 
expansion {\it i.e. } $B_{\rm eff}\cos(\sqrt{2K}\phi_+)\to B_{\rm eff}K\phi_+^2$.
The calculation of $F_{-1}^+$ becomes then similar to the one performed in Ref. \onlinecite{starykh}. 
The result is (see also Refs. \onlinecite{voit:1998} and \onlinecite{wiegmann:1999})
\begin{equation}\label{eq:resf+}
	F_{-1}^+(x,\tau)
	\approx 
	\e^{-\frac{1}{K} \left[\frac{1}{2K_+^*}\ln\left(\frac{x^2+v_+^2\tau^2}{a^2}\right)+\frac{m}{2K_+^*}\sqrt{x^2+v_+^2\tau^2} \right]},
\end{equation}
where $m^2=B_{\rm eff}K/a v_+$, 
$K=K_c+1/K_s$ as defined before, and $K_+^*$ is the strong coupling LL constant of the 
``$+$'' sector of the model defined by Eq. \eqref{eq:H_pm}.
Contrary to Ref. \onlinecite{starykh}, $K_+^*$ is only little renormalized by the renormalization group flow of the 
cosine term in Eq. \eqref{eq:H_pm},\cite{bsl2} and we can use $K_+^* \approx 1$.

Since the quadratic expansion of the cosine term in  Eq. (\ref{eq:H_pm}) may seem a rather crude
approximation, one can check that Eq. (\ref{eq:resf+}) makes sense by comparing it to the non-interacting result obtained 
for $K_c=K_s=1$. In this limit the cosine term in Eq. (\ref{eq:H_pm}) is equal to $\cos(2\phi_+)$, and corresponds to the 
backscattering of $R\dw$ and $L\up$ fermions (the ones associated with the ``$+$'' sector). 
Moreover $F_{-1}^{+,ni}(x,\tau)$ is 
proportional to the Green's function of these massive fermions and therefore can be written as
\be
	F_{-1}^{+,ni}(x,\tau)
	=
	2\pi a\int \frac{d\omega_n dq}{(2\pi)^2} 
	\e^{i\omega_n\tau+i q x}\frac{-i\omega_n-vq}{\omega_n^2+\Delta^2+v^2q^2},
\ee
with $\Delta\propto \sqrt{B_{\rm eff} v/a}$
and $v=v_+=v_-$.
At $x=0$, we can evaluate $F_{-1}^{+,ni}(0,\tau)$ analytically and find
\be  
	F_{-1}^{+,ni}(0,\tau)=\frac{\Delta}{2v}K_1(\Delta|\tau|),
\ee
where $K_1$ is the modified Bessel function.
Since $K_1(z)\sim \e^{-z}/\sqrt{z}$ for $z\gg 1$, the asymptotic behavior of
$F_{-1}^+(0,\tau)$ in the non-interacting limit is consistent with the behavior described in Eq. (\ref{eq:resf+}). 
This indicates that the rather crude approximation used to calculate $F_{-1}^+(x,\tau)$ for a given $K$ captures 
correctly the asymptotic limit for the Green's function.

The full $(x,i\omega_n)$ dependence of $F_{-1}^{+}$ cannot be determined analytically. We can, however, 
give a solution in terms of a convolution integral, similar to Eq. \eqref{eq:G_conv}, if we write
$F_{-1}^+(x,\tau)= f_i(x,\tau) f_m(x,\tau)$, with $f_i$ the irregular part from the gapless sector,
$f_i(x,\tau) = [a^2/(x^2+v_+^2\tau^2)]^{1/2K}$, and $f_m$ the massive part, 
$f_m(x,\tau) = \e^{- \frac{m}{2K} \sqrt{x^2+v_+^2 \tau^2}}$.
The gapless part $f_i(x,i\omega_n)$ is identical to $I(x,i\omega_n)$ in Eq. \eqref{eq:I} with 
the replacements $\gamma \to 1/2K$ and $v \to v_+$ and the solution given by Eq. \eqref{eq:I_sol}. 
The $\tau$ integral for the 
massive part $f_m(x,i\omega_n)$ can be evaluated through standard integral tables (Ref. \onlinecite{gradshteyn}, 3.914)
and we find
\begin{align}
	f_m(x,i\omega_n)	
	&= 2 \int_0^\infty d\tau \, \cos(\omega_n \tau) \, \e^{- \frac{m}{2K} \sqrt{x^2+v_+^2 \tau^2}}
\nonumber \\
	&= \frac{m |x|}{v_+ \beta} K_1\left(|x| \beta / v_+\right),
\end{align}
with $\beta = \sqrt{\frac{v_+^2 m^2}{4 K^2} + \omega_n^2}$.
Using the latter expressions, 
\begin{equation} \label{eq:F^+_{-1}_convolution}
	F^+_{-1}(x,i\omega_n) = \int d\omega_n' f_i(x,i\omega_n') f_m(x,i\omega_n-i\omega_n'),
\end{equation}
which is again well suited for numerical evaluation because of the exponential 
convergence of the Bessel function $K_1$.

The second factor for the Green's function in Eq. \eqref{eq:gr},
$F_{-1}^-(x,\tau)$, is entirely evaluated within the gapless LL sector 
and is given by
\begin{equation}\label{eq:F_1}
F_{-1}^-(x,\tau)
	= - \sign(\tau)
	\left[\frac{-i a}{x - i v_- \tau}\right]^{\gamma_{irr}}
	\left[\frac{i a}{x + i v_- \tau}\right]^{\gamma_{irr}},
\end{equation}
with
\begin{equation}
	\gamma_{irr} = \frac{\bar{K}^2}{4 K_{SLL} K^2}
	= \frac{\bar{K}^2 K_s}{8 K_c K}
	= \frac{K_s (K_c-K_s^{-1})^2}{8 K_c (K_c+K_s^{-1})}.
\end{equation}
The $\sign(\tau)$ factor in Eq. \eqref{eq:F_1} is necessary to ensure
the correct fermion	 statistics, which is  no longer enforced by the
form of the power-law factor, since it depends now only on $x^2 + v_-^2 \tau^2$
(the sign can also be verified by an analytic continuation of the 
independently calculated real time Green's function).
In contrast to the regular LL, both exponents in Eq. \eqref{eq:F_1}
are identical, expressing
the fact that the left- and right-movers are pinned to each other
[see Eq. \eqref{eq:phi_Rdw_phi_Lup}].

The Fourier transform  resembles
closely the integral $I(x,i\omega_n)$ of Eq. \eqref{eq:I}, with the supplementary factor
$\sign(\tau)$ in the integrand. 
The result can again be evaluated using standard integral tables
[Ref. \onlinecite{gradshteyn}, 3.771.1, Ref. \onlinecite{abramowitz}, 12.2.3].
It is less compact than for $I(x,i\omega_n)$ and given by 
\begin{align}
	&F_{-1}^{-}(x,i \omega_n)
	=
	-i \sign(\omega_n) \frac{a\sqrt{\pi}\Gamma(1-\gamma_{irr})}{2  v_-}
	\left| \frac{2 \omega_n a^2}{xv_-}\right|^{\gamma_{irr}-\frac{1}{2} }
\nonumber \\
&\times
	\left[ I_{\gamma_{irr}-\frac{1}{2}}(|x \omega_n|/v_-) - \mathbf{L}_{\frac{1}{2}-\gamma_{irr}}(|x \omega_n|/v_-) \right],
\label{eq:G_irr_Mats}
\end{align}
where $\mathbf{L}_\alpha(z)$ is the modified Struve function and $I_\alpha(z)$ the modified Bessel function.

We again perform an independent calculation for the retarded Green's function associated with $F_{-1}^-$, 
denoted by $F_{-1}^{-,ret}$, for which the sign $\sign(\tau) \to \sign(t)$ leads to the further terms
described at the end of Appendix \ref{ap:retarded}. 
The result is
\begin{align}
	&F_{-1}^{-,ret}(x,\omega)
	=
	-\sign(\omega)
	\frac{ 2a\sqrt{\pi}}{v_-}
	\Biggl\{
		\frac{1}{\Gamma(\gamma_{irr})}
		\left(\frac{2i |x|v_-}{\omega_+a^2}\right)^{\frac{1}{2}-\gamma_{irr}}
\nonumber \\
&\quad\times
		K_{\frac{1}{2}-\gamma_{irr}}(|x|\omega_+ /iv_-)
	-
		\frac{\Gamma(1-\gamma_{irr})}{2}
		\left|\frac{2x v_-}{\omega a^2 }\right|^{\frac{1}{2}-\gamma_{irr}}
\nonumber \\
&\quad\times
		\left[
			\mathbf{H}_{\frac{1}{2}-\gamma_{irr}}(|\omega x|/v_-) - Y_{\frac{1}{2}-\gamma_{irr}}(|\omega x|/v_-)
		\right]
	\Biggr\},
\label{eq:G_irr_ret_xomega}
\end{align}
where $\mathbf{H}_\alpha(x)$ denotes the Struve function,  $Y_\alpha(x)$
the Bessel function of the second kind, and $\omega_+ = \omega + i 0$.
The part depending on the Bessel function $K_{\frac{1}{2}-\gamma_{irr}}$ corresponds
to $I(x,\omega)$ given in Eq. \eqref{eq:I_ret_sol}, while the other
terms are parts of the integral that cancel for $I(x,\omega)$ but contribute here
because of the additional factor $\sign(t)$. They are given by Eq. \eqref{eq:I_vert}.
We notice, however, that the latter terms
are all purely real so mainly the complex $K_{\frac{1}{2}-\gamma_{irr}}$ dependent term
contributes to the spectral properties of the system.
The properties of the Struve functions are described in Ref. \onlinecite{abramowitz}, Sec. 12.
The Struve functions are related to the modified Struve functions by 
$\mathbf{L}_\alpha(z) = -i \e^{-i \frac{\pi}{2}\alpha} \mathbf{H}_\alpha(i z)$.
The asymptotic behavior of the Green's functions at large $z$ is determined by noting that 
$\mathbf{H}_{\frac{1}{2}-\gamma}(z) - Y_{\frac{1}{2}-\gamma}(z) \sim \mathcal{O}(z^{-\frac{1}{2}-\gamma})$
and
$I_{\gamma-\frac{1}{2}}(z) - \mathbf{L}_{\frac{1}{2}-\gamma}(z) \sim \mathcal{O}(z^{-\frac{1}{2}-\gamma})$,
so Eq. \eqref{eq:G_irr_Mats}, as well as the last terms in Eq. \eqref{eq:G_irr_ret_xomega} have the large-distance 
non-oscillating decays $|x|^{-1} |\omega|^{-1}$ and $|x|^{-1} |\omega|^{-2\gamma_{irr}}$, respectively.
For $z \to 0$, the Struve functions go to zero, while $I_{\gamma-\frac{1}{2}}$ and $Y_{\frac{1}{2}-\gamma}$ have 
divergences that are regularized by the ultraviolet cutoff.

The convolution of the $F^\pm_{-1}(x,i\omega_n)$ times the prefactor $-\e^{-irk_F x}/2\pi a$ yields 
the Green's function $\tilde{G}_{r\sigma,r\sigma}(x,\tau)$.

Let us, finally, note that cross-correlators $\bar{r} \neq \bar{r}'$ in the irregular sector, 
with $\bar{r},\bar{r}' \in \{ \mbox{$(L\up)$}, \mbox{$(R\dw)$} \}$, are nonzero.
The corresponding Green's functions $\mean{T_\tau \psi_{\bar{r}}(x,\tau) \psi_{\bar{r}'}^\dagger(x',\tau')}$ actually depend on the two 
position variables $x$ and $x'$. This is because, for the cross-correlators, the response is no longer translationally 
invariant due to the phase factors $\e^{\pm i k_F (x+x')}$. This spatial variation is the only difference between the auto-correlators
and the cross-correlators. Hence the result for the cross-correlators is identical to the result of the 
auto-correlators determined above, on the replacement $x \to x-x'$ everywhere, except in the 
phase factor $\e^{ i r k_F x}$, where we make the change $x \to x+x'$.


\section{Density of states}
\label{sec:dos}

The DOS per unit volume $a$ (or tunneling DOS) $\rho(\omega)$,
can be obtained from the imaginary
part of the retarded local Green's function as
\be
	\rho(\omega) = -\frac{1}{\pi}{\rm Im} G^{ret}(x\to 0,\omega),
\ee
where $G^{ret} = \sum\limits_{r,\sigma} G^{ret}_{r\sigma,r\sigma}$.

\subsection{Density of states for the Luttinger liquid and the Helical Luttinger liquid}\label{sec:dosll}

The calculation of the DOS for a spinless LL is detailed in Appendix \ref{ap:alpha}.
Starting from Eq. \eqref{eq:G_ret_LL} one obtains
\be \label{eq:dos_ll}
	\rho(\omega) = 2 \frac{|\omega/\Delta_a|^{2 \gamma}}{2\pi v\Gamma(1+2\gamma)},
\ee
with $\gamma = (K+K^{-1}-2)/4$ and $\Delta_a = v/a$ on the order of the bandwidth.
The prefactor 2 arises from the sum over $r=L,R$.
Here we have dropped a divergent, frequency independent contribution to the DOS. 
The latter is a well-known artifact from the LL theory in the short distance
limit $x \to 0$, and must in practice be limited by the value of the short-distance cutoff $a$.
The form of the DOS in Eq.~\eqref{eq:dos_ll} corresponds to the results known in the literature.\cite{gogolin,giamarchi}
For non-interacting electrons, $\gamma \to 0$ and the DOS becomes a constant, reflecting
the fact that a linear dispersion is assumed in LL theory.

On the replacement $K \to K_{HLL}$ in the expression of $\gamma$, the result \eqref{eq:dos_ll} holds as well for the HLL.

For a spinful LL, we have noticed in Sec. \ref{sec:Greens_LL} that the $x\to 0$
limit can be evaluated using the Green's functions of a spinless LL while replacing
$\gamma \to \gamma_c + \gamma_s = (K_c + K_c^{-1} + K_s + K_s^{-1} - 4)/8$ and
$v\to \sqrt{v_c v_s}$. Since $v_{c,s} = v_F / K_{c,s}$, we can incorporate this rescaling of
velocities by redefining $\Delta_a = \sqrt{v_c v_s} / a = v_F / a \sqrt{K_c K_s}$.
This yields
\be \label{eq:dos_ll_spin}
	\rho(\omega) =
	4 \frac{|\omega/\Delta_a|^{2(\gamma_c+\gamma_s)}}{2\pi v \Gamma(1+2(\gamma_c+\gamma_s))},
\ee
where the four channels $(r,\sigma)$ contribute identically to the DOS.

\subsection{Density of states for a Spiral Luttinger liquid}

For a SLL, the DOS is the sum of the contributions from the regular and irregular sectors. 
The spin-resolved DOS is
\begin{equation} \label{eq:rho_LL_spin}
	\rho_\sigma(\omega)
	=
	-\frac{1}{\pi}{\rm Im}[G^{ret}_{L\sigma} (x\to 0,\omega)+G^{ret}_{R\sigma} (x\to 0,\omega)].
\end{equation}
Independently of $\sigma$, one of the two contributions in Eq.~\eqref{eq:rho_LL_spin} arises from the regular sector,
and the other one from the irregular sector, therefore
\begin{equation}
	\rho_\sigma(\omega)
	= \rho_{reg}(\omega) + \rho_{irr}(\omega).
\end{equation}
For comparison, we consider, first, the non-interacting limit, since in the absence of interactions simple expressions can be 
obtained for the DOS. In this limit $\gamma_{irr}=\gamma_{reg}=0$, 
and the DOS associated with the irregular sector does not 
couple to the gapless fields. It is given by the DOS of the gapped system only,
\be
	\rho_{irr}^{ni}(\omega,x=0)=\frac{1}{2\pi v}\vartheta(|\omega|-\Delta)\frac{|\omega|}{\sqrt{\omega^2-\Delta^2}}.
	\label{ni}
\ee
Since $\rho_{reg}^{ni}(\omega,x=0)=1/2\pi v_F$, the total DOS exhibits a pseudo-gap for 
$|\omega|<\Delta$ corresponding to the pinning of half of the degrees of freedom.

Let us now consider the general interacting case. In Sec. \ref{sec:regular_sector_GF} we have seen 
that the Green's function of the regular sector coincides with the Green's function of
a spinless LL with the substitutions $v \to v_{SLL}$ and $\gamma \to \gamma_{reg}$.
Hence, the corresponding DOS can be read off from Eq. \eqref{eq:dos_ll} and is given by 
\be\label{eq:rho_reg}
	\rho_{reg}(\omega)=
	\frac{|\omega/\Delta_a^{-}|^{2 \gamma_{reg}}}{2\pi v_{SLL} \Gamma(1+2\gamma_{reg})},
\ee
where $\Delta_a^{-}=v_{SLL}/a = v_-/a$. Note that the prefactor $2$ in Eq. \eqref{eq:dos_ll} is 
absent because $\rho_{reg}$ refers here to a single mode $(L \dw)$ or $(R \up)$.
 
The irregular part of the DOS can, following Ref.~\onlinecite{starykh}, be expressed by
\be\label{eq:rho_integral}
	\rho_{irr}(\omega)
	=
	\frac{1}{\pi^3 a}
	\int_0^\omega d\epsilon  
	\mathrm{Im}[F_{-1}^{+,ret}(0,\epsilon)]
	\mathrm{Im}[F_{-1}^{-,ret}(0,\omega-\epsilon)],
\ee
which is proved in Appendix \ref{ap:rho_integral}.
The function $F_{-1}^{-,ret}(x,\omega)$ is given by Eq. (\ref{eq:G_irr_ret_xomega}). 
The only part that is not purely real and contributes to $\mathrm{Im}[F_{-1}^{-,ret}(0,\omega-\epsilon)]$ is
the part depending on the Bessel function $K_{\frac{1}{2}-\gamma_{irr}}$. 
Expanding the Bessel function for small 
$x$ (see Appendix \ref{ap:alpha}) leads to 
\be
	\mathrm{Im}[F_{-1}^{-,ret}(0,\omega)]
	=
	- \frac{\pi}{\Delta_a^{-} \Gamma(2\gamma_{irr})} \left| \frac{\omega}{\Delta_a^{-}} \right|^{2\gamma_{irr}-1}.
\ee
The function $F_{-1}^{+}(x,i\omega_n)$ was given in Eq. \eqref{eq:F^+_{-1}_convolution} in form of a convolution 
integral and so is, in principle, more difficult to evaluate. However, 
for the DOS only the $x\to 0$ limit is required, for which
we can solve the Fourier integral for $F^+_{-1}(x=0,\omega_n)$ directly. 
Indeed, at $x=0$, the Fourier transformation of Eq. \eqref{eq:resf+} reduces 
to the standard integral of the Gamma function and we obtain
\begin{align}
	&F_{-1}^+(x=0,i\omega_n)
	= 
	\left(\frac{a}{v_+}\right)^{\frac{1}{K}}
	\Gamma\left(1-\frac{1}{K}\right)
\nonumber \\
&\times
	\left[
		\left(\frac{m v_+}{2K}-i\omega_n\right)^{\frac{1}{K}-1}
		+
		\left(\frac{m v_+}{2K}+i\omega_n\right)^{\frac{1}{K}-1}
	\right].
\label{eq:F_-1^+_x=0}
\end{align}
With the analytic continuation $i\omega_n \to \omega_+$ we then find
\begin{align}
	\mathrm{Im}[F_{-1}^{+,ret}(0,\omega)]
	&= 
	-
	\sign(\omega)
	\vartheta\left(|\omega|-\Delta \right)
\nonumber \\
&\times
	\frac{\pi}{\Delta_a^{+}\Gamma\left(\frac{1}{K}\right)}
	\left(\frac{|\omega|-\Delta }{\Delta_a^+}\right)^{\frac{1}{K}-1},
\end{align}
where we have used $\sin(\frac{\pi}{K}) \Gamma(1-\frac{1}{K}) = \pi/\Gamma(\frac{1}{K})$
and defined
$\Delta_a^+ = v_+/a$ as well as
$\Delta = mv_+/2K = \sqrt{B_{\rm eff}\Delta_a^+/4K}$.
From the step function $\vartheta$ in the latter equation follows that $\rho_{irr}(\omega) = 0$ 
for $|\omega| < \Delta$. For $\omega > \Delta$ we then have
\begin{align}
	\rho_{irr}(\omega)
	&=
	\frac{1}{\pi a (\Delta_a^{-})^{2\gamma_{irr}} (\Delta_a^+)^{\frac{1}{K}} \Gamma(2\gamma_{irr})\Gamma(\frac{1}{K})}
\nonumber \\
	&\times
	\int_{\Delta}^\omega 
	d\epsilon (\omega-\epsilon)^{2\gamma_{irr}-1} (\epsilon-\Delta)^{\frac{1}{K}-1},
\nonumber \\
	&=
	\frac{1}{\pi a (\Delta_a^{-})^{2\gamma_{irr}} (\Delta_a^+)^{\frac{1}{K}} \Gamma(2\gamma_{irr})\Gamma(\frac{1}{K})}
\nonumber \\
	&\times
	(\omega-\Delta)^{2\gamma_{irr}+\frac{1}{K}-1}
	B\left(2\gamma_{irr}, \frac{1}{K}\right),
\end{align}
with $B$ the Beta function.
This result can be simplified by using
$2\gamma_{irr}+\frac{1}{K} = \frac{1}{2 K_{SLL}}$ and $B(x,y)=\Gamma(x)\Gamma(y)/\Gamma(x+y)$.
Extending it furthermore to all $\omega$ we find
\begin{align}
	\rho_{irr}(\omega)
	&=
	\frac{\vartheta(|\omega|-\Delta)}{\Gamma(\frac{1}{2K_{SLL}})}	
	\frac{(|\omega|-\Delta)^{\frac{1}{2K_{SLL}}-1}}
	     {\pi a (\Delta_a^{-})^{2\gamma_{irr}} (\Delta_a^{+})^{\frac{1}{K}}}.
\label{eq:rho_irr}
\end{align}
Note that even though $\rho_{irr}$ contains the gapless part 
$\mathrm{Im}F^-_{-1}$ the threshold behavior of the massive sector $\propto \vartheta(|\omega|-\Delta)$
is preserved in $\rho_{irr}$. The power-law at $|\omega| > \Delta$, however, is the combination of
the massive and the gapless contributions and curiously turns out to be given by the LL parameter 
$K_{SLL}$ of the gapless fields $\phi_-,\theta_-$ only,
which is a behavior found also for the DOS of the gapped sector in charge density wave systems.\cite{schuricht:2011a}

In the non-interacting limit, $K_{SLL} \to 1$ and $\gamma_{irr} \to 0$, 
the frequency dependence of $\rho_{irr}^{ni}$ in the vicinity of $|\omega| \sim \Delta$ is recovered. 
However, there is a difference to Eq. \eqref{ni} by a numerical prefactor of order 1. This is
an artifact from the high-energy cutoff procedures and just reflects 
the unavoidable uncertainties with numerical prefactors in LL theory.

The total DOS is subsequently obtained as the sum 
$\rho(\omega) = \rho_\up(\omega) + \rho_\dw(\omega) = 2[\rho_{reg}(\omega)+\rho_{irr}(\omega)]$, 
where the 2 arises from summing over the two regular $\{(L\dw),(R\up)\}$ as well as the two
irregular $\{(L\up),(R\dw)\}$ contributions to the DOS.

Let us note that since cross-correlators in the irregular sector between $(L\up)$ and
$(R\dw)$ modes are nonzero, the spin-polarization densities of state $\sigma_{x}(\omega)$ or $\sigma_{y}(\omega)$
probe directly $\rho_{irr}$, as they are linear combinations of 
$\mean{\psi_\up^\dagger \psi_\dw}$ and $\mean{\psi_\dw^\dagger \psi_\up}$. 
This is a consequence of the partial electron ordering in form of a spiral electron spin 
density wave, pinned to the nuclear spin spiral, or of the transverse ferromagnetic electron order aligned with the 
external magnetic field for the quantum wires with spin-orbit interaction at commensurability.

\begin{figure}
	\begin{center}
	\includegraphics[width=0.9\columnwidth]{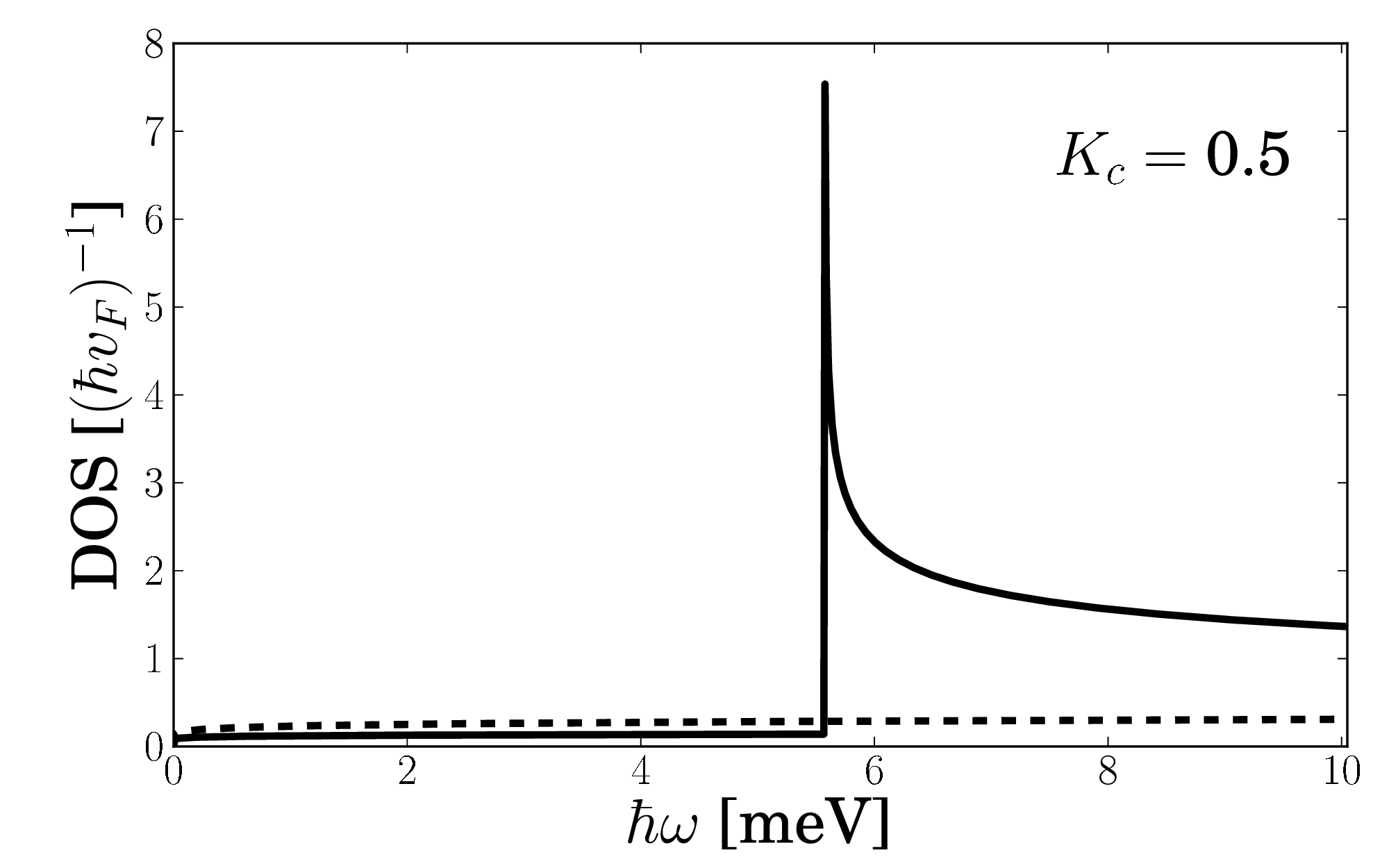}
	\end{center}
	\caption{The local DOS as a function of energy for the SLL appearing through 
	nuclear spin order in GaAs quantum wires (solid line). 
	The LL parameters are chosen as\cite{bsl2}
	$K_c = 0.5$ and $K_s = 1$. Further system parameters as well as the size
	of the pseudogap $\Delta$ (after renormalization by electron interactions)
	are determined according to Ref. \onlinecite{bsl2}.
	Below $\Delta$, the DOS is described only by $\rho_{reg}$, while at
	$\omega\gtrsim\Delta$ the DOS is dominated by the singular behavior of 
	$\rho_{irr}$. For comparison, the dashed line shows the DOS of a regular
	spinful LL with the same parameters $K_c$ and $K_s$.
	\label{fig:dos1}}
\end{figure}

\begin{figure}
	\begin{center}
	\includegraphics[width=0.9\columnwidth]{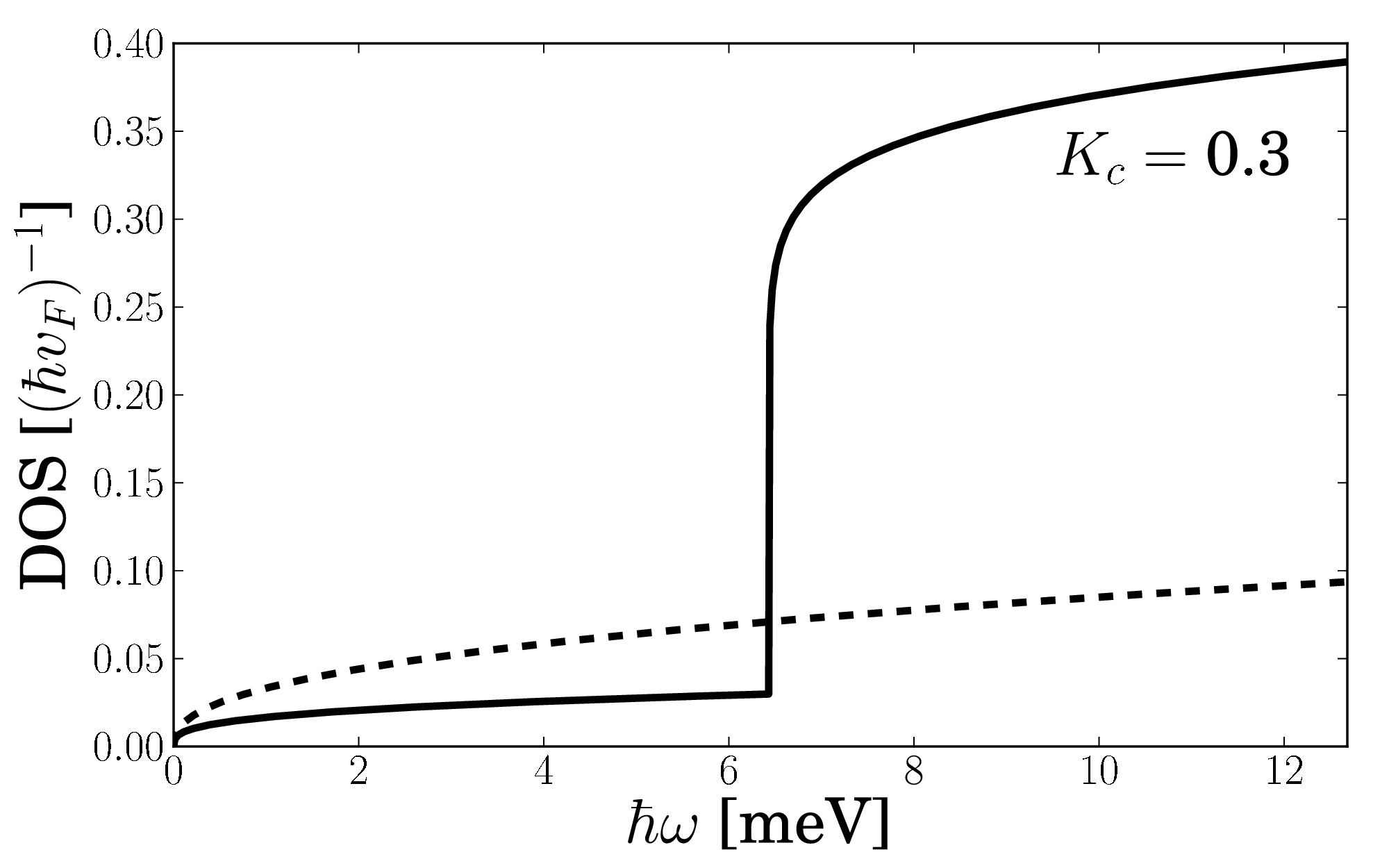}
	\end{center}
	\caption{\label{fig:dos2}
	Same as in Fig. \ref{fig:dos1}, but for the reduced value $K_c = 0.3$.
	Since $K_c < 1/3$, the
	power-law singularity of $\rho_{irr}$ at $\omega=\Delta$ is replaced by 
	$\rho_{irr} \to 0$. Yet, $\rho_{irr}$ still dominates the DOS 
	at $\omega \gtrsim \Delta$. Note that $\Delta$ takes here a different
	value than in Fig. \ref{fig:dos1} because of the different renormalization by
	the stronger interactions.\cite{bsl2}}
\end{figure}

In Figs. \ref{fig:dos1} and \ref{fig:dos2} we plot the DOS as a function of energy
for the SLL appearing through nuclear spin ordering in typical GaAs quantum wires\cite{bsl2}
for $K_s=1$ and $K_c > 1/3$ as well as $K_c < 1/3$. For $K_s = 1$ the value $K_c=1/3$ 
marks the point at which $K_{SLL} = 1/2$, and at which 
the exponent in Eq. \eqref{eq:rho_irr} changes its sign. For $K_c > 1/3$ the irregular
contribution diverges at $|\omega| = \Delta$, while for $K_c < 1/3$ it goes to zero.
In both cases, however, Figs. \ref{fig:dos1} and \ref{fig:dos2} show that $\rho_{irr}$
dominates the DOS at $|\omega| \gtrsim \Delta$. 
A similar effect is observed when 
calculating the tunneling current between a LL and a superconductor,\cite{stmpaper} which is not surprising, since both 
calculations involve convolutions between a LL-type power-law decaying component, and a gapped component.
It needs to be stressed furthermore that our analysis is valid only in the vicinity 
of $\Delta$, as further fluctuations of the gapped fields are not taken into account.
However, as shown in a very recent complementary study of the SLL,\cite{schuricht:2011c}
the inclusion of such fluctuations does not change the overall behavior of the DOS
close to $\Delta$ but leads to modified power-law exponents. 

%

\section{Conclusions}
\label{sec:conclusions}

In this paper we have computed the position-frequency dependent Green's functions
and the local DOS for various LLs. Most of the results are given in closed analytical
form, to our knowledge, for the first time. These Green's functions are necessary for 
any local spectral characterization of the LLs, and, in particular, we have used them
to derive analytic expressions for the local DOS of all the investigated LLs. 
While the latter is a well-known result for the standard LLs,\cite{gogolin,giamarchi}
we have put special attention to the HLL appearing at the edge of topological insulators
and the SLL that appears in systems that are often used to mimic HLL behavior. We have shown
that while HLL and SLL are indeed equivalent for non-interacting electrons, they have
substantially different spectral properties for interacting electrons. While the HLL
behaves as a standard spinless LL, the SLL shows with the appearance of the irregular
parts of the Green's functions entirely different response functions. The main reason 
for this striking difference is that the SLL exhibits the response of a standard gapless
LL together with that of a gapped system with a dynamically created gap. The response
therefore exhibits a pseudogap. 
Close to the pseudogap, the strong fluctuations of the fields conjugate
to the gapped fields lead to the strong irregular response that dominates the 
spectral properties. Below the pseudogap, the local response described by the DOS
shows a regular LL behavior, yet with modified exponents. 
Interestingly, the irregular behavior has an effect even below the pseudogap
when considering the nonlocal response described by the Green's functions at $x \neq 0$.
This behavior is entirely absent in any HLL. Hence any analysis of the system properties
and application that is based on the apparent equivalence of HLL and SLL needs to take
this important difference into account.

Concerning the detection of the discussed SLL properties it should be stressed that the SLL physics
arises only when the chemical potential is kept near the center of the pseudogap. 
This occurs naturally in the nuclear-spin-ordered phase, in which the gap opens symmetrically
about the chemical potential. It requires, however, a fine tuning of the electron density to $k_F = k_{so}$
for the spin-orbit split quantum wires. This means, it is not possible to probe SLL physics
by modulating the chemical potential, and techniques such as STM need to be used.

The systems that can exhibit the SLL physics are 1D conductors in
the nuclear ordered phase such as GaAs quantum wires\cite{bsl2} or carbon nanotubes\cite{bsl1,bsl2} 
grown entirely from $^{13}$C isotopes to provide the nuclear spins.\cite{simon,ruemmeli,churchill_1,churchill_2}
In these
systems, the particular spectral properties of the SLL could be used to 
indirectly prove the existence of the nuclear spin order. Furthermore, SLL
physics is expected to arise in 1D conductors with strong spin-orbit
interaction such as GaAs quantum wires, InAs nanowires, or Ge/Si core/shell nanowires
in the presence of an external magnetic field,\cite{streda:2003,pershin:2004,devillard:2005,zhang:2006,sanchez:2008,birkholz:2009,quay:2010,bjkl,lu:2005,xiang:2006,park:2010,lee:2010,kloeffel:2011,hao:2010}
as well as carbon nanotubes in strong electric fields.\cite{ksbl1,ksbl2}


\begin{acknowledgments} 
We thank S. Gangadharaiah, D. L. Maslov, D. Loss and D. Schuricht for useful discussions, 
and we thank M. Hohenadler and R. Lutchyn for helpful correspondence.
B.B. acknowledges the support by the EU-FP7 project SE2ND [271554], the Swiss SNF, NCCR Nanoscience (Basel), and NCCR QSIT.
C.B. is supported by the 
ERC Starting Independent Researcher Grant NANO-GRAPHENE 256965 and in part by the National Science Foundation under 
Grant No. 1066293, and acknowledges the hospitality of the Aspen Center for Physics.
P.S. acknowledges the kind hospitality of the Department of Physics of the University of Basel where part of this work was completed.
\end{acknowledgments}


\appendix


\section{Retarded Green's functions in $(x,\omega)$ space for the regular spinless Luttinger liquid}
\label{ap:retarded}

In this appendix we provide the details of the calculation of the
real-time Fourier transform of the retarded Green's functions.

\begin{figure}
	\begin{center}
		\includegraphics[width=\columnwidth]{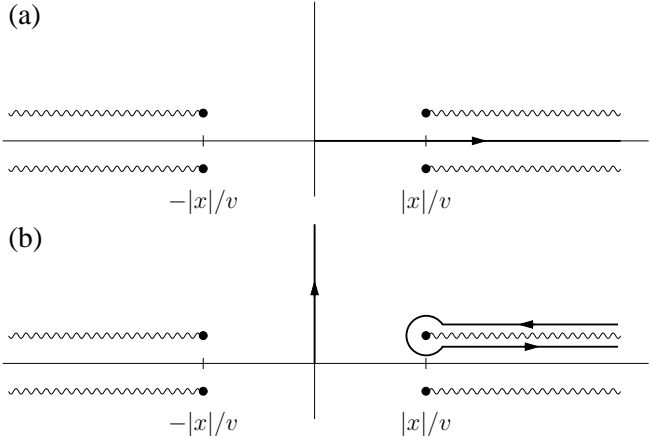}
	\end{center}
	\caption{%
	Branch cuts and integration contours for Eq. \eqref{eq:I_ret}
	in the complex $t$ plane.
	(a) The first term in Eq. \eqref{eq:I_ret} has the upper right and lower
	left branch cuts, and the second term the upper left and lower right branch cuts.
	The integration
	runs along the real axis from $t=0$ to $\infty$.
	(b) Since $\omega > 0$, the integrand converges in the upper half
	plane and can be deformed to a part running along the positive
	imaginary axis, and a part running around the upper right branch cut.
	Along the imaginary axis, both terms in the integrand are identical
	and cancel each other. The contribution from the circular contour around
	the branch point at $|x|/v + i0$ vanishes. The two legs above and below
	the branch cut give Eq. \eqref{eq:I_ret_int}, where $-2i\sin(\pi \gamma)$
	captures the phase difference between the upper and lower legs.
	\label{fig:contours}}
\end{figure}

The greater and lesser functions can be
inferred directly from the imaginary time Green's function.
Indeed, since the latter is given by power-laws in $x \pm iv \tau$,
the real time Green's functions can be obtained by a simple analytic
continuation of the \emph{time} variable,\cite{giamarchi,negele}
$\tilde{G}(x,\tau \to i (t-i0)) = -i G^>(x,t)$, and
$\tilde{G}(x,\tau \to i (t+i0)) = i G^<(x,t)$.
Due to the power-law forms, this substitution is equivalent to the
usual Wick rotation in frequency space and is valid as long as the
analytical continuation does not lie on a branch cut.
 From Eq. \eqref{eq:G_tau0} we have
\begin{align}
	&G^{>,<}_{r,r}(x,t)
\nonumber\\
	&= \mp i\frac{\e^{i r k_F x}}{2\pi a}
	\left[\frac{-i a}{x + v (t\mp i0)}\right]^{\gamma_r}
	\left[\frac{i a}{x - v (t\mp i0)}\right]^{\gamma'_r}
\nonumber\\
	&= \frac{\e^{i r k_F x}}{2\pi a}
	\left[\frac{a}{rx + v (t\mp i0)}\right]^{\gamma}
	\left[\frac{a}{rx - v (t\mp i0)}\right]^{\gamma+1},
\label{eq:G><_xt}
\end{align}
with $\gamma = \min(\gamma_L,\gamma_R) = (K+K^{-1}-2)/4$.
Since $\gamma+1 > 1$, the singularity at $rx-vt=0$ requires regularization,
which can be done in the same way as in Eq. \eqref{eq:G_tau} by
reducing the exponent to $\gamma$ by derivation with respect to
$\frac{-1}{2\gamma}(\partial_{rx} - \partial_{v t})$.
We apply this directly on the retarded Green's function
$G^{ret}_{r,r}(x,t) = \vartheta(t) [ G^>_{r,r}(x,t)-G^<_{r,r}]$, and
obtain for the Fourier transform
\begin{equation}
	G_{r,r}^{ret}(x,\omega)
	= \frac{-\e^{i r k_F x}}{4\pi \gamma}
	\left[ \partial_{rx} + \frac{i \omega_+}{v} \right]
	I(x,\omega),
\label{eq:G^ret_I}
\end{equation}
with $\omega_+ = \omega + i 0$ and
\begin{align}
	&I(x,\omega)
	=
	\int_0^\infty dt \, \e^{i \omega_+ t}
\nonumber \\
	&\times
	\left\{
		\left[\frac{a}{x^2 + v^2(t-i0)^2}\right]^\gamma
		-
		\left[\frac{a}{x^2 + v^2(t+i0)^2}\right]^\gamma
	\right\}.
\label{eq:I_ret}
\end{align}
The term depending on $(t-i0)$ has branch cuts starting at
$\pm |x|/v +i 0$ running to $\pm \infty$ slightly above the real axis,
and the term depending on $(t+i0)$ has the same branch cuts shifted
slightly below the real axis [see Fig. \ref{fig:contours} (a)].
Since $\omega$ is real, we have $I(x,-\omega) = - I^*(x,\omega)$,
allowing us to focus on $\omega > 0$.
We can then deform the integration contour as shown in Fig. \ref{fig:contours} (b).
The only nonzero contribution comes from the two straight lines above and
below the branch cut from $|x|/v$ to $+\infty$. This leads to
(with $y = t + |x|/v$)
\begin{equation} \label{eq:I_ret_int}
	I(x,\omega)
	= \frac{-2 i \sin(\pi \gamma)}{(v/a)^{2\gamma}} \e^{i \omega_+ |x|/v}
	\int_{0}^\infty dy \, \frac{\e^{i \omega_+ y}}{[y(y+2|x|/v)]^\gamma}.
\end{equation}
Due to the $+i0$ in $\omega_+$ this integral is convergent and
very similar to Eq. \eqref{eq:I}.
In fact, if the integration contour in Eq. \eqref{eq:I} is deformed such that it runs
around a branch cut, the integral becomes identical to Eq. \eqref{eq:I_ret_int}
under the replacement $|\omega_n| \to -i \omega_+$.
This integral can be evaluated using Ref. \onlinecite{gradshteyn}, 3.383.8, and produces
a result equivalent to Eq. \eqref{eq:I_sol},
\begin{equation} \label{eq:I_ret_sol}
	I(x,\omega)
	=
	-i \frac{2a\sqrt{\pi}}{v\Gamma(\gamma)}
	\left(\frac{2 i |x| v}{\omega_+ a^2}\right)^{\frac{1}{2}-\gamma} K_{\gamma-\frac{1}{2}}(|x| \omega_+/iv).
\end{equation}
This form also holds for $\omega < 0$ as it satisfies $I(x,\omega) = - I^*(x,-\omega)$.
The branch as a function of $\omega$ runs along the negative imaginary axis.
Therefore, the Wick rotation $\omega_+ \to i \omega_n$ is indeed
problematic for $\omega_n < 0$, which explains why separate calculations were necessary
for the real and imaginary time Green's functions.

The final formula for the retarded Green's function is obtained in the same way as in
Sec. \ref{sec:GF} using Eq. \eqref{eq:K_id}. We find
\begin{align}
	&G^{ret}_{r,r}(x,\omega)
	= \frac{-\e^{i r k_F x} }{2\sqrt{\pi} \Gamma(\gamma+1)} \frac{a\omega_+}{v^2}
	\left(\frac{2i|x|v}{\omega_+ a^2}\right)^{\frac{1}{2}-\gamma}
\nonumber \\
	&\times
	\left[ K_{\gamma-\frac{1}{2}}(|x| \omega_+/iv) - \sign(r x) K_{\gamma+\frac{1}{2}}(|x|\omega_+/iv) \right],
\end{align}
which is the result reported in Eq. \eqref{eq:G_ret_LL}.

To finish this section, let us calculate the contribution of the vertical contour on the positive imaginary
axis in Fig. \ref{fig:contours} (b). While this integral has dropped out for $I(x,\omega)$, it 
actually contributes to the Green's function of the irregular sector of Eq. \eqref{eq:F_1}, 
which is of the form of Eq. \eqref{eq:I_ret}, yet with a $+$ sign between the two power laws.
We denote this contribution by $I^\text{vert}$.
Using the standard table, Ref. \onlinecite{gradshteyn}, 3.387.7, we see that it evaluates to
\begin{align}
	&I^{\text{vert}}(x,\omega)
	=
	2i \int_0^\infty dt \e^{-|\omega| t} \left[\frac{a^2}{x^2+v^2 t^2}\right]^{\gamma}
\nonumber \\
	&= 
	i \frac{a \sqrt{\pi}}{v} 
	\Gamma(1-\gamma)
	\left|\frac{2 x v}{\omega a^2}\right|^{\frac{1}{2}-\gamma}
\nonumber \\
	&\times
	\left[
		\mathbf{H}_{\frac{1}{2}-\gamma}\left(\frac{|x\omega|}{v}\right) 
		- 
		Y_{\frac{1}{2}-\gamma}\left(\frac{|x\omega|}{v}\right)
	\right],
	\label{eq:I_vert}
\end{align}
with $\mathbf{H}_\alpha$ the Struve function  and $Y_\alpha$ the Bessel function of
the second kind. This function, in combination with the solution $I(x,\omega)$ of Eq. \eqref{eq:I_ret_sol},
provides the result of Eq. \eqref{eq:G_irr_ret_xomega}.


\section{Details for the calculation of the density of states}
\label{ap:alpha}

We evaluate in this appendix the limit $x \to 0$ for the retarded Green's function
required for the DOS $\rho(\omega)$.
Letting $x \to 0$ and rescaling Eq. \eqref{eq:G><_xt} or \eqref{eq:G_tau0} by $t \to \omega t$
we see that $\rho(\omega) \propto \omega^{2g}$ for the regular LL. However, we see also that
taking the $x \to 0$ limit before evaluating the Fourier integral over $t$ makes the integral divergent.
Taking the limit $x \to 0$ after integration does not cure this problem, and the power-law divergences
that arise from the extrapolation of the long-wavelength LL theory into
the short distance $x \to 0$ regime are still present. This is, however, an artifact of the lack of small 
distance cutoff in the model and should have no influence on the DOS; thus, these terms can be safely 
neglected. In fact, they can be entirely suppressed by the regularization via a further differentiation 
of the integrand as in Eq. \eqref{eq:G_tau}.

The goal of the present appendix is to evaluate the prefactors of the $\omega^{2\gamma}$
terms for a typical LL.  This analysis is based on an expansion of
the Bessel functions $K_\alpha$ at small argument (see Ref. \onlinecite{abramowitz}, 9.6.9),
\begin{equation} \label{eq:K_nu_expansion}
	K_\alpha(z) \overset{z \to 0}{\sim}
	\frac{1}{2}
	\left[
		\Gamma(\alpha)  \Bigl(\frac{2}{z}\Bigr)^{\alpha}
		+
		\Gamma(-\alpha) \Bigl(\frac{z}{2}\Bigr)^{\alpha}
	\right]
	\bigl(1+\mathcal{O}(z^2)\bigr).
\end{equation}
Introducing this expansion into Eq. \eqref{eq:G_ret_LL} for a regular spinless LL
and neglecting the divergent parts, we obtain
\begin{align}
	G^{ret}_{r,r}(x,\omega)
	\overset{x\to 0}{\sim}
	\frac{-\Gamma\left(\frac{1}{2}-\gamma\right)}{4 \sqrt{\pi} \Gamma(1+\gamma)}
	\frac{\omega_+ a}{v^2}
	\left(\frac{\omega_+ a}{2iv}\right)^{2\gamma-1},
\end{align}
so
\begin{align}
	\rho_r(\omega)
	&=
	- \frac{1}{\pi} \text{Im} G^{ret}_{r,r}(x\to 0,\omega)
\nonumber \\
	&=
	\frac{1}{2\pi v}
	\frac{\cos(\pi\gamma) \Gamma(\frac{1}{2}-\gamma)}{2^{2\gamma}\sqrt{\pi} \Gamma(\gamma+1)}
	\left|\frac{\omega a}{v}\right|^{2\gamma}.	
\end{align}
Using the doubling theorem of Gamma functions [Ref. \onlinecite{abramowitz}, 6.1.18],
$\Gamma(2z) = (2\pi)^{-1/2}  2^{2z-\frac{1}{2}} \Gamma(z) \Gamma(z+\frac{1}{2})$
with $z = \frac{1}{2}-\gamma$, we can
further simplify this result to
\begin{equation}
	\rho_r(\omega)
	=
	\frac{1}{2\pi v}
	\frac{1}{\Gamma(1+2\gamma)}
	\left|\frac{\omega a}{v}\right|^{2\gamma}.
\end{equation}

\section{Proof of Eq. (\ref{eq:rho_integral})}
\label{ap:rho_integral}

\begin{figure}
	\begin{center}
		\includegraphics[width=0.6\columnwidth]{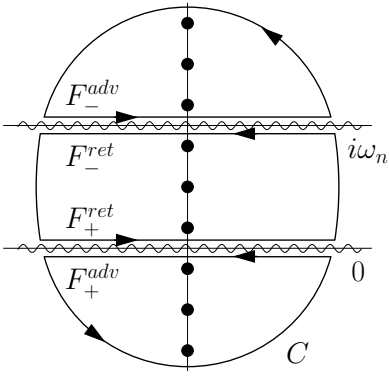}
	\end{center}
	\caption{%
	Integration contour $C$ for Eq. \eqref{eq:G_contour_C}. The black dots
	indicate the Matsubara poles. The wiggled lines mark the branch cuts
	at $\mathrm{Im}(z)=0$ separating retarded and advanced parts of $F_+$,
	and at $\mathrm{Im}(z)=i\omega_n$ separating retarded and advanced parts of $F_-$.
	\label{fig:contour_im}}
\end{figure}

In this appendix we derive the identity given by Eq. \eqref{eq:rho_integral}.
To this end, it is useful to change to a finite temperature description with 
discrete Matsubara frequencies. In this case, the Green's function of Eq. \eqref{eq:gr}
has the generic form
\begin{equation}
	\tilde{G}(i\omega_n)
	=
	\sum_{i\omega_m} F_+(i\omega_m) F_-(i\omega_n-i\omega_m),
\end{equation} 
where we omit all prefactors and indices that are not necessary for the 
current calculation. 
We are here confronted with an ambiguity of how to choose the Matsubara frequencies
in the bosonization formulation, as $i\omega_n$ should in principle be a fermionic 
frequency. Independently of the choice of $i\omega_m$ one of the $F_\pm$ would then
depend on a bosonic and the other one on a fermionic Matsubara frequency. In the limit
of zero temperature, however, all Matsubara frequencies become continuous, and so we
make here the most convenient choice of bosonic $i\omega_n = 2 n \pi/\beta$ and 
fermionic $i\omega_m = (2m+1)\pi/ \beta$, for integers $n$ and $m$, and $\beta$ the inverse
temperature. This choice guarantees an equivalent treatment of both $F_\pm$ and allows
us to avoid unnecessary complications with poles on the branch cuts.
In a standard way, we can replace the Matsubara sum by the integral
\begin{equation} \label{eq:G_contour_C}
	\tilde{G}(i\omega_n)
	= 
	\int_C \frac{dz}{2\pi i} \tanh\left(\frac{\beta z}{2}\right) F_+(z) F_-(i\omega_n-z),
\end{equation}
with $C$ a contour surrounding all Matsubara frequencies $i\omega_m$. 
Since $F_+(z)$
has a branch cut at $\mathrm{Im}(z) = 0$ and $F_-(i\omega_n-z)$
has a branch cut at $\mathrm{Im}(z) = i\omega_n$, the contour $C$ takes the form
shown in Fig. \ref{fig:contour_im}. 
Each branch cut separates the retarded and advanced
parts of the corresponding $F_\pm$. Since $i\omega_n$ is bosonic, there is no pole on 
the branch cuts, and the contour integral reduces to the evaluation along the cuts,
\begin{align}
	&\tilde{G}(i\omega_n)
	= 
	\int_{-\infty}^\infty \frac{d\omega'}{2\pi i} 
	\tanh\left(\frac{\beta \omega'}{2}\right)
	\left[ F_+^{ret}(\omega')-F_+^{adv}(\omega') \right] 
\nonumber\\
	&\times
	F_-^{ret}(i\omega_n-\omega')
	+\int_{-\infty}^\infty \frac{d\omega'}{2\pi i} 
	\tanh\left(\frac{\beta (i\omega_n+\omega')}{2}\right)
\nonumber\\
	&\times
	F_+^{ret}(i\omega_n+\omega')
	\left[ F_-^{adv}(-\omega')-F_-^{ret}(-\omega') \right].
\end{align}
We now perform the analytic continuation $i\omega_n \to \omega_+ = \omega+i0$.
Here we need to emphasize that the analytic continuation is ill defined for the 
$\tanh$, as it is periodic in $i\omega_n$. To maintain the uniqueness of the 
analytic continuation, we must first drop the $i\omega_n$ dependence of the 
$\tanh$ and then perform the analytic continuation for the functions $F_\pm$ only.
The imaginary part of the resulting retarded Green's function 
can then be written as (after shifting the integration variable in the 
second integral by $\omega$),
\begin{align}
	&\mathrm{Im}G^{ret}(\omega)
\nonumber\\
	&=
	\int_{-\infty}^\infty \frac{d\omega'}{\pi} 
	\left[
		\tanh\left(\frac{\beta \omega'}{2}\right)
		-	
		\tanh\left(\frac{\beta (\omega'-\omega)}{2}\right)
	\right]	
\nonumber\\
	&\quad\times
	\mathrm{Im}[F_+^{ret}(\omega')] 
	\mathrm{Im}[F_-^{ret}(\omega-\omega')].
\end{align}
At zero temperature $\beta \to \infty$ and the difference of $\tanh$ 
restricts the integration to the interval $(0,\omega)$ for $\omega>0$ 
or $(\omega,0)$ for $\omega<0$, and provides a 
factor $2 \sign(\omega)$. The result is 
\begin{align}
	\mathrm{Im}G^{ret}(\omega)
	=
	\frac{2}{\pi} \int_{0}^\omega d\omega' 
	\mathrm{Im}[F_+^{ret}(\omega')] 
	\mathrm{Im}[F_-^{ret}(\omega-\omega')],
\end{align}
which corresponds to Eq. \eqref{eq:rho_integral}.


\end{document}